\documentclass[aps,prl,amsmath,amssymb,twocolumn,twoside,showpacs,superscriptaddress]{revtex4-2}
\usepackage[pdftex]{graphicx}
\usepackage[colorlinks=true,citecolor=blue]{hyperref}
\usepackage{graphicx,textcomp,dcolumn,hyperref,upgreek}

\usepackage{physics}

\usepackage{dsfont}

\newcommand{\z}{_{\noindent z}}

\begin{document}

\title{
Pushing the limits  in real-time measurements of quantum dynamics
}
\author{E.~Kleinherbers}
\email{eric.kleinherbers@uni-due.de}
\affiliation{Faculty of Physics and CENIDE, University of Duisburg-Essen, 47057 Duisburg, Germany}
\author{P.~Stegmann}
\email{psteg@mit.edu}
\affiliation{Department of Chemistry, Massachusetts Institute of Technology, Cambridge, Massachusetts 02139, USA}
\author{A.~Kurzmann}
\affiliation{2nd Institute of Physics, RWTH Aachen University, 52074 Aachen, Germany}
\author{M.~Geller}
\affiliation{Faculty of Physics and CENIDE, University of Duisburg-Essen, 47057 Duisburg, Germany}
\author{A.~Lorke}
\affiliation{Faculty of Physics and CENIDE, University of Duisburg-Essen, 47057 Duisburg, Germany}
\author{J.~K{\"o}nig}
\affiliation{Faculty of Physics and CENIDE, University of Duisburg-Essen, 47057 Duisburg, Germany}
               
\date{\today}

\begin{abstract}
Time-resolved studies of quantum systems are the key to understand quantum dynamics at its core. The real-time measurement of individual quantum numbers as they switch between certain discrete values, well known as \textit{random telegraph signal}, is expected to yield maximal physical insight. 
However, the signal suffers from both systematic errors, such as a limited time resolution and noise from the measurement apparatus, as well as statistical errors due to a limited amount of data.
Here we demonstrate that an evaluation scheme based on factorial cumulants can reduce the influence of such errors by orders of magnitude.
The error resilience is supported by a general theory for the detection errors as well as experimental data of single-electron tunneling through a self-assembled quantum dot. 
Thus, factorial cumulants push the limits in the analysis of random telegraph data which represent a wide class of experiments in physics, chemistry, engineering and life sciences.
\end{abstract}

\maketitle

Resolving dynamics of open quantum systems~\cite{breuer_theory_2002} on the most fundamental level of individual quantum events is a common goal in many fields of science.
Real-time measurements have been performed for a large variety of quantum systems, including ions~\cite{hume_2011}, neutral atoms~\cite{natterer2017,saskin_narrow_2019}, single molecules~\cite{zhou2011detecting,choi2012single,chung2012single,xin2019concepts,miyamachi_2012,burzuri2018spin}, and skyrmions~\cite{muckel_2021}.
Fluctuating occupation numbers of optical and plasmonic cavities~\cite{guerlin_progressive_2007,gupta_2021}, metallic islands~\cite{pekola_single_2013}, quantum dots~\cite{efros_1997,gustavsson2006counting,flindt2009universal,komijani2013counting,matsuo_2020}, trapped quantum gases~\cite{Ott_single_2016}, and nanocalorimeters~\cite{karimi_2020_nat,karimi_2020} have been measured with single-photon, -electron, -atom, and -ion precision.

\begin{figure}[t]
	\includegraphics[width=8.6cm]{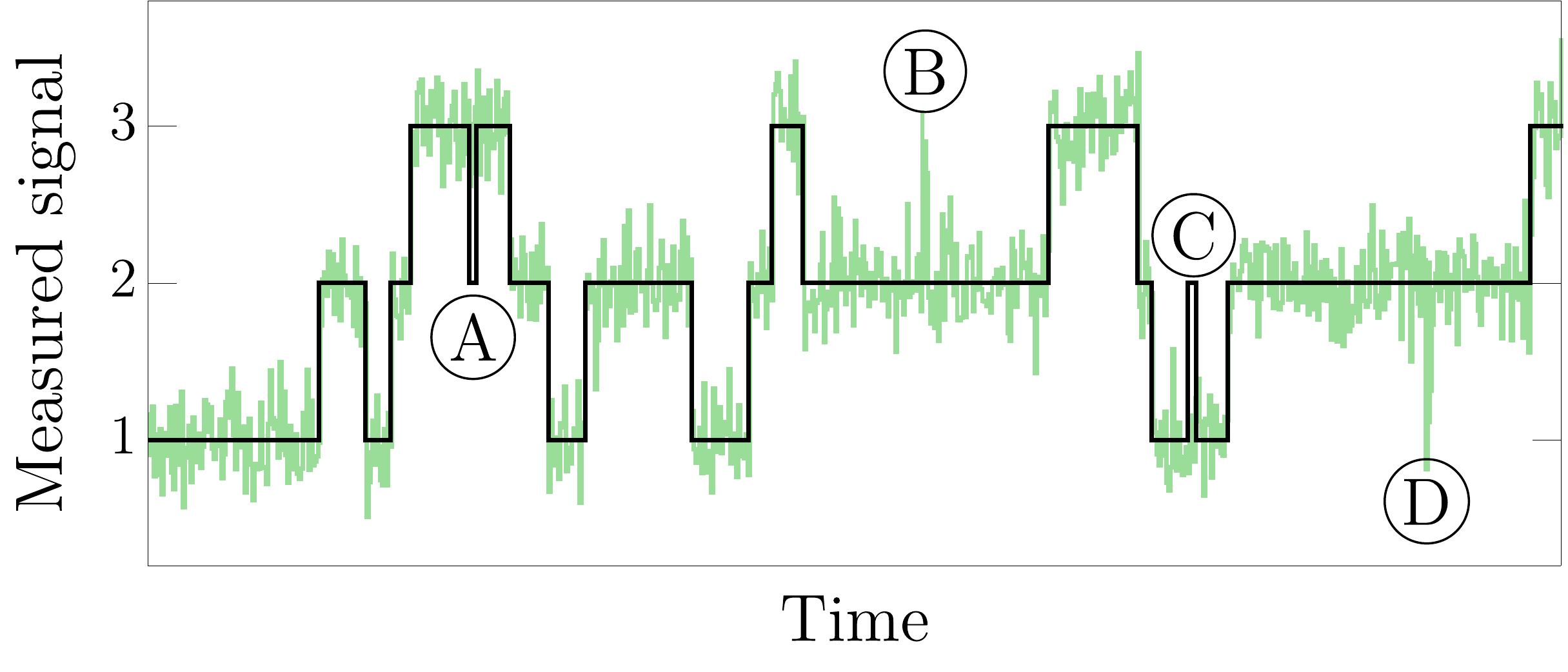}
	\caption{Generic form of a random telegraph signal (green) that deviates from the true quantum dynamics (black) because of events that are missed (A and C) or falsely indicated (B and D) by the detector. Simulated data is depicted.
}
\label{fig:1}	
\end{figure}

These experiments record in time switches between distinct quantum states, as illustrated by the black line in Fig.~\ref{fig:1}. 
The form of the depicted time evolution is known as random telegraph signal.
It can provide information about hidden quantum states such as degenerate spin states~\cite{matsuo_2020} or dark states~\cite{gupta_2021}. Underlying interactions such as magnetic~\cite{natterer2017,saskin_narrow_2019} or  attractive electron-electron interactions~\cite{kleinherbers2018revealing} as well as internal quantum transitions such as spin relaxation~\cite{kurzmann2019optical} or conformational changes in single molecules~\cite{zhou2011detecting,choi2012single,chung2012single,xin2019concepts,miyamachi_2012,burzuri2018spin} can be revealed. 
Unfortunately, the measured signal (green line) suffers from problems that appear in any detection scheme:
fast transitions are overlooked (indicated by A and C) due to a limited time resolution, false transitions (indicated by B and D) are recorded due to a noisy detector signal, and statistical errors occur due to the finite time span over which data is collected.

There are many experimental attempts to overcome these problems,
e.g. by employing high-bandwidth detection~\cite{kurzmann2019optical}, 
noise-suppression techniques~\cite{prechtel_2013,hansom_2014,wagner_2017,al2018photon}, 
or quantum stochastic resonance~\cite{wagner2019quantum}.
As a complementary strategy to push the limits set by typical detection errors, we suggest
to employ a specific statistical tool-set, i.e. \textit{factorial cumulants}, for the analysis of random telegraph data.
Factorial cumulants are well known from a mathematical perspective~\cite{johnson_univariate_2005} and designed to characterize discrete probability distributions~\cite{koenig_2021}, in contrast to ordinary cumulants which are designed for continuous probability distributions. 
Therefore, it is much more natural to use factorial cumulants for the analysis of random telegraph data, where transitions between discrete states are investigated.
Moreover, factorial cumulants are expected to be sensitive indicators for correlation~\cite{beenakker_counting_2001,kambly2011factorial,stegmann2015detection,stegmann2016short,kleinherbers2018revealing}. 
Nonetheless, their full potential has so far only been little explored for practical evaluation of (noisy) statistical data. 
This is very unfortunate since, as we show in this Letter, factorial cumulants are resilient to errors that otherwise obscure the quantum dynamics of interest and may result in a wrong modeling of the quantum system.

\begin{figure}[t]
	\includegraphics[width=8.6cm]{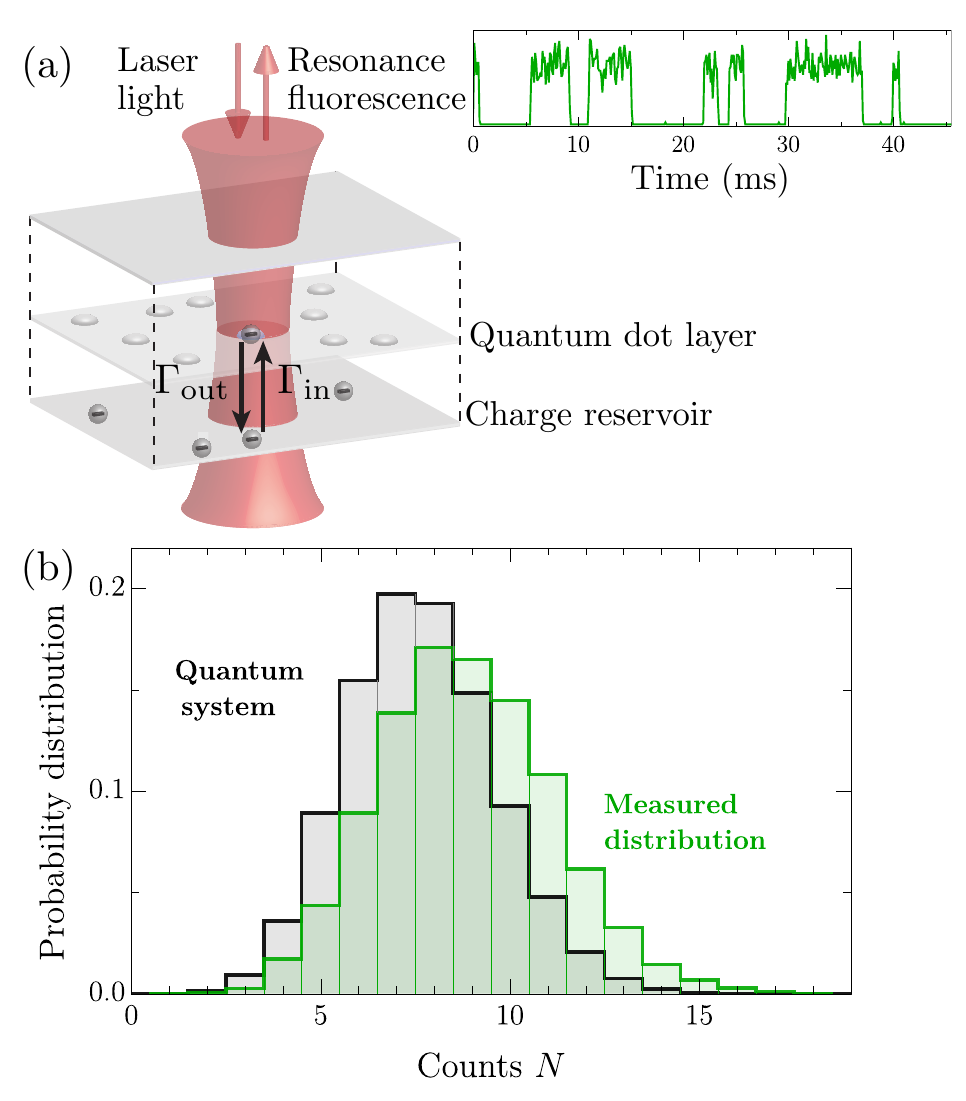}
	\caption{(a) Experimental setup for the optical readout of the electron occupation of a self-assembled quantum dot. The measured resonance fluorescence signal is depicted in green. (b) Measured probability distribution $P_N^\text{meas}$ (green) compared with the distribution $P_N$ (black) that originates from the true quantum dynamics for a time interval $t=45\,\text{ms}$.
}
\label{fig:2}	
\end{figure}

To illustrate this concept, we study temporal charge fluctuations of a self-assembled semiconductor quantum dot with single-electron precision. 
The setup is depicted in Fig.~\ref{fig:2}(a). The quantum dot is tunnel coupled to an external charge reservoir, so that single electrons can tunnel into and out of the quantum dot with rates $\Gamma_\text{in}$ and $\Gamma_\text{out}$, respectively. 
Due to a strong Coulomb repulsion, the quantum dot is either empty or occupied by one electron only.
The occupation of the quantum dot is monitored using a resonance-fluorescence readout scheme~\cite{vamivakas_2010,kurzmann_2016,kurzmann2019optical,lochner_2020}. If the quantum dot is empty, an infrared laser drives an excitonic transition and the emitted fluorescence photons are collected by a single-photon detector. If the quantum dot is occupied, no photons are emitted. 
After binning the measured stream of single photons with an adjustable binning time, the bright state (empty quantum dot) and the dark state (occupied quantum dot) can be resolved as a function of time, see the resonance fluorescence signal in Fig.~\ref{fig:2}(a) depicted in green.
The setup yields a high-quality telegraph signal with almost negligible errors as a reference measurement. Nevertheless, we can artificially increase the detection errors by either modifying the time resolution or removing a large fraction of the collected photons and thus ``blinding'' the detector.

\begin{figure}[t]
	\includegraphics[width=8.6cm]{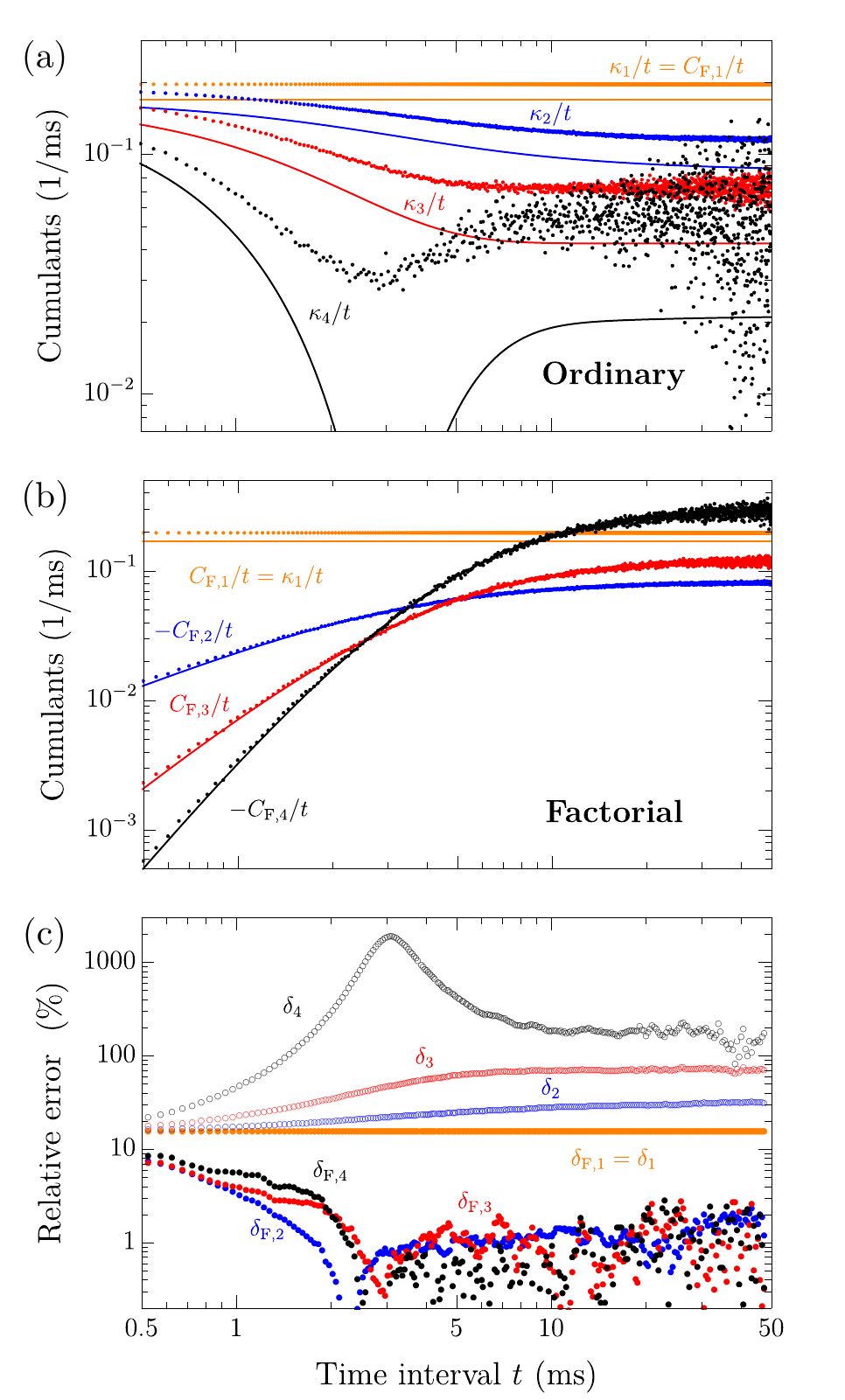}
	\caption{(a) Ordinary cumulants $\kappa_m/t$, (b) factorial cumulants $C_{\text{F},m}/t$, and (c) their relative errors $\delta_m$ and $\delta_{\text{F},m}$ as a function of time $t$. Experimental data (dots) is compared with a simulation disregarding measurement imperfections (solid lines). Relative errors in (c) are obtained by averaging over 20 successive data points to reduce statistical errors. The time resolution is $\Delta t=50\,\mu\text{s}$, the false-count rate of the bright state is $\Gamma_0^\text{false}=0.059\,\text{kHz}$, and the duration of the measurement is $T=369\,\text{s}$. The electron-tunneling rates are $\Gamma_\text{in}=0.346\,\text{kHz}$ and $\Gamma_\text{out}=0.334\,\text{kHz}$.} 
\label{fig:3}
\end{figure}

The measured telegraph signal contains much more information than just the mean number $\ev{N}$ of tunneling events. In particular, the fluctuations around this mean value have a strong predictive power about the properties of the quantum system~\cite{blanter_2000}. 
In the framework of full counting statistics, the information of these fluctuations is summarized in the probability distribution $P_N^\text{meas}(t)$ that $N$ tunneling events have been counted in a time interval of length $t$~[green histogram in Fig.~\ref{fig:2}(b)], where we use the convention to count only tunneling-out events. 

The measured probability distribution $P_N^\text{meas}$ can be systematically analyzed by its ordinary cumulants $\kappa_m$ of order $m$~\cite{gustavsson_electron_2009}. The first cumulant  $\kappa_1=\ev{N}$ describes the mean and the second cumulant $\kappa_2={\ev{N^2}}{-}\ev{N}^2$ the variance of the distribution. 
With increasing order $m$, successively more details about $P_N^\text{meas}$ are revealed. The cumulants can be derived from the generating function
\begin{align}\label{eq:1new}
S^\text{meas}(z)= \text{ln} \Bigl(\sum_N z^N P^\text{meas}_N \Bigr),
\end{align}
via $\kappa_m= \partial_\chi^m S^\text{meas}(e^\chi)\vert_{\chi=0}$~\cite{gustavsson_electron_2009}, where we introduce the counting variable~$z$. 
In Fig.~\ref{fig:3}(a), the ordinary cumulants $\kappa_m$ (dots) are 
depicted as a function of time $t$. 
As the order $m$ increases, the time dependence $\kappa_m(t)$ acquires more and more structure. However, this is merely part of a general property of ordinary cumulants, referred to as universal oscillations~\cite{flindt2009universal}, and hence contains no system-specific information. Therefore, it has been suggested to use factorial cumulants $C_{\text{F},m}(t)$ instead~\cite{kambly2011factorial}, which are defined by $C_{\text{F},m}= \partial_z^m S^\text{meas}(z)\vert_{z=1}$.
They are related to ordinary cumulants by $C_{\text{F},m}=\sum_{j=1}^m s_{m,j} \kappa_j$, with the Stirling numbers of the first kind $s_{m,j}$ giving the coefficients of the factorial power~\cite{johnson_univariate_2005}. 
In fact, in Fig.~\ref{fig:3}(b), the factorial cumulants $C_{\text{F},m}$ (dots) do not show    such universal oscillations and, thus, are much better suited to extract physical information. 

In this Letter, we demonstrate an even more remarkable advantage of factorial cumulants, namely their robustness against errors, which is also clearly visible in Fig.~\ref{fig:3}(a),(b). 
We compare the measured cumulants (dots) with the theoretical limit (solid lines) of an ideal measurement with infinite bandwidth and signal-to-noise ratio, as well as an unlimited amount of data. While the ordinary cumulants $\kappa_m(t)$ in Fig.~\ref{fig:3}(a) are heavily influenced by the measurement imperfections, the factorial cumulants  $C_{\text{F},m}(t)$  in Fig.~\ref{fig:3}(b), on the other hand, are error resilient.  
The relative error depicted in Fig.~\ref{fig:3}(c) increases drastically for the ordinary cumulants $\kappa_m(t)$ with each order $m$ and surpasses $100\%$ beginning with the fourth ordinary cumulant at finite times. In contrast, the error of the factorial cumulants $C_{\text{F},m}(t)$ at finite times remains at around $1 \%$ for all orders $m>1$.

\begin{figure}[t]
	\includegraphics[width=8.cm]{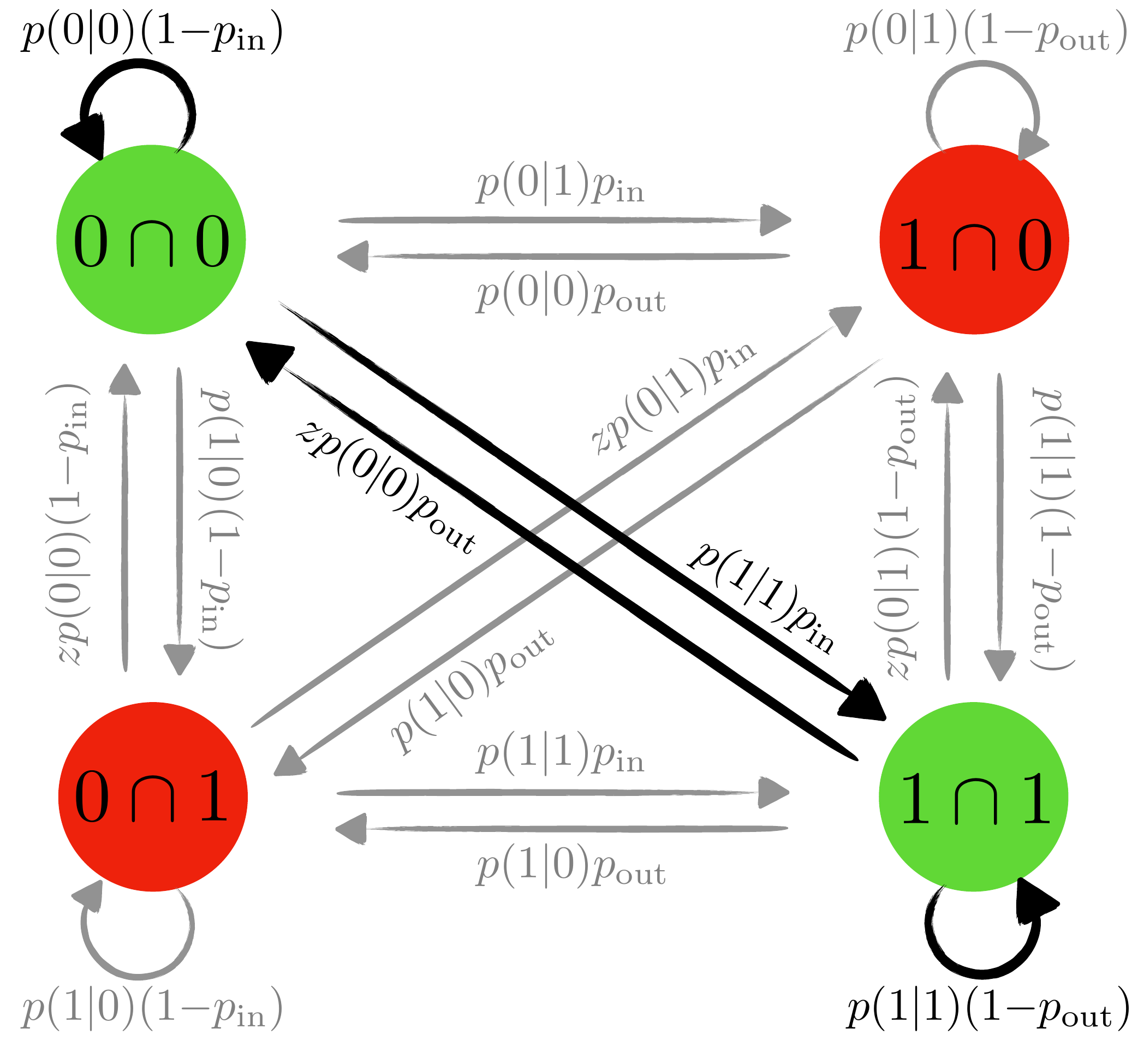}
	\caption{Four dimensional model to simulate both a limited time resolution and noise. The possible states are indicated via $a\cap b$ denoting that the measurement outcome is $b$ and the true value is $a$. True associations $b=a$ are colored in green and false associations $b\neq a$ are colored in red.  Each time step $\Delta t$, the states are updated due to true tunneling events with transition probabilities $p_\text{in}$ and $p_\text{out}$ and false noise-induced events with probabilities $p(b\vert a)$. Noise-related transitions are indicated as gray arrows. Transitions increasing the detector counter $N$ are multiplied by $z$.}
\label{fig:4}	
\end{figure}

To explain the ruggedness of factorial cumulants against measurement imperfections, we need to faithfully model the measured probability distribution $P_N^\text{meas}$. 
In any detection scheme, the probability distribution is inevitably subjected to errors and, thus, can be decomposed as
\begin{align}\label{eq:2new}
P_N^\text{meas}=\sum_{N^\prime=0}^N P_{N{-}N^\prime} \, \delta P^\text{sys}_{N^\prime}+\delta P^{\text{sta}}_N.
\end{align}
The desired information about the electron tunneling events is contained in $P_N$~[black histogram in Fig.~\ref{fig:2}(b)]. In contrast, $\delta P^\text{sys}_{N}$ accounts for the systematic error due to missed and false events, and $\delta P^\text{sta}_N$ represents the statistical error caused by the finite measurement time.
Accordingly, we can write the generating function from Eq.~\eqref{eq:1new} as
\begin{align}\label{eq:3new}
S^\text{meas}=S+\delta S^\text{sys}+\delta S^\text{sta}.
\end{align}
The function $S=\ln (\sum{z^N P_N})$ is related via $P_N=\tr[{\rho}_N(t)]$ to the quantum system's density matrix $\rho_N$ with the constrain that $N$ tunneling-out events have occurred in the time interval $[0,t]$.
The time evolution of $\rho_N$ is governed by the $N$-resolved master equation~\cite{plenio_1998,flindt_2005}
\begin{align}\label{eq:4new}
\dot{\rho}_N=({\cal W}{-}{\cal J}_{\text{out}}) \rho_{N}+ {\cal J}_{\text{out}} \rho_{N-1},
\end{align}
where ${\cal W}$ is the generator of the full time evolution, while ${\cal J}_{\text{out}}={\cal P}_0{\cal W}{\cal P}_{1}$ describes the tunneling-out events from the occupied (projector ${\cal P}_{1}$) to the empty (projector ${\cal P}_0$) quantum dot.
The solution of the master equation is readily obtained after a $z$-transform and reads $\rho\z=\sum_N z^N \rho_N= e^{{\cal W}\z t}\rho_\text{st}$ with the generator ${\cal W}\z=({\cal W}-{\cal J}_{\text{out}})+z {\cal J}_{\text{out}}$. The stationary state of the quantum system $\rho_\text{st}$ has been reached before the counting starts. Finally, tracing out the quantum degrees of freedom leads to the generating function~\cite{breuer_theory_2002,flindt_2005}
\begin{align}\label{eq:5new}
S= \ln \tr( e^{{\cal W}\z t} \rho_\text{st}).
\end{align} 

A unified theoretical description of the errors $\delta S^\text{sys}$ and $\delta S^{\text{sta}}$ has been missing in the literature so far. 
However, neglecting them may result in a huge discrepancy between experiment and (error-free) theoretical model, as illustrated in Fig.~\ref{fig:3}(a).
To close this gap, we developed a general model that accounts for measurement imperfections and can be applied to an arbitrary quantum system and an arbitrary set of detected and undetected quantum transitions~(Supplementary Section~I). 
Using the quantum dot system as an example, we present here the steps to incorporate the errors into the theoretical model which is illustrated in Fig.~\ref{fig:4}.

First, we take into account that the quantum dot state is measured with a limited time resolution $\Delta t$.
Therefore, the counter $N$ is not introduced on the level of the master equation~\eqref{eq:4new}, but on the level of the coarse-grained time evolution
\begin{align}\label{eq:6new}
\rho_N(t+\Delta t)=(\Pi{-}{\cal P}_0 \Pi {\cal P}_1) \rho_{N}(t)+ {\cal P}_0 \Pi {\cal P}_1 \rho_{N-1}(t),
\end{align}
which ensures that each time step $\Delta t$ the counter $N$ increases at most by one.  Here, $\Pi = e^{{\cal W} \Delta t}$ propagates the quantum state in steps of $\Delta t$. Transitions from the empty (0) to the occupied dot (1) and vice versa happen each time step $\Delta t$ with probability $p_\text{in}$ and $p_\text{out}$, respectively. With probability $1{-}p_\text{in}$ and $1{-}p_\text{out}$ the state does not change.

Second, to account for a faulty detector, whose output may deviate from the actual quantum state, we explicitly introduce the detector degree of freedom. Therefore, we resolve the density matrix according to $\boldsymbol{\varrho}_N=(\rho_N^{(0)},\rho_N^{(1)})$, where the superscript $(b)$ with $b\in\{0,1\}$ denotes the state indicated by the detector.
Thus, the density matrix element $\mel{a}{\rho^{(b)}}{a}=p(a\cap b)$ gives the joint probability that the detector output is $b$ and the quantum dot state is $\ket{a}$ with $a\in\{0,1\}$. In Fig.~\ref{fig:4}, true associations $a=b$ are shown in green and false associations  $a\neq b$ are shown in red. The $N$-resolved time evolution becomes
\begin{align}\label{eq:7new}
\boldsymbol{\varrho}_N(t+\Delta t)=(\boldsymbol{\Pi}{-}{\cal P}^{(0)} \boldsymbol{\Pi} {\cal P}^{(1)}) \boldsymbol{\varrho}_{N}(t)+ {\cal P}^{(0)} \boldsymbol{\Pi} {\cal P}^{(1)}\boldsymbol{\varrho}_{N-1}(t),
\end{align}
where the projectors ${\cal P}^{(b)}$ for the detector states ensure that the counter $N$ is only sensitive to changes of the detector output. The propagator is given by $\boldsymbol{\Pi}= {\cal F}\!\cdot\!( {\cal D} \otimes  e^{{\cal W} \Delta t})$ with $({\cal D})_{bb^\prime}=1$. Here, the diagonal matrix ${\cal F}=\text{diag}\big[p(0\vert 0),p(0\vert 1),p(1\vert 0),p(1\vert 1)\big]$ accounts for false detector outputs, where $p(b\vert a)$ are the conditional probabilities that we measure $b$, given that the true value is $a$. They fulfill $\sum_b p(b\vert a)=1$. Thus, each time step $\Delta t$, the detector indicates with a probability $p(0\vert 1)$ an empty and with $p(1\vert 0)$ an occupied quantum dot although the actual state is the opposite, see Fig.~\ref{fig:4}. False transitions of the form $0\rightarrow1$ and $1\rightarrow0$ (similar to B and D in Fig.~\ref{fig:1}) are the consequence. The type of the noise and $\Delta t$ determine the specific values of the conditional probabilities $p(b\vert a)$.

To solve Eq.~\eqref{eq:7new}, we perform a $z$-transform and find
\begin{align}\label{eq:8new}
\boldsymbol{\varrho}\z (t+\Delta t)=\boldsymbol{\Pi}\z\boldsymbol{\varrho}\z(t)= {\cal F} \cdot \big( {{\cal D}\z} \otimes e^{{\cal W}\Delta t} \big) \boldsymbol{\varrho}\z(t)
\end{align}
with $({\cal D}\z)_{bb^\prime}=1+(z-1)\delta_{b0}\delta_{1b^\prime}$. Starting from the stationary state $\boldsymbol{\varrho}_\text{st}$, we apply Eq.~\eqref{eq:8new} successively to arrive at $\boldsymbol{\varrho}\z (t)=\boldsymbol{\Pi}\z ^{t/\Delta t}\boldsymbol{\varrho}_\text{st}$. Finally, we trace out the quantum degrees of freedom and obtain the full generating function
\begin{align}\label{eq:9new}
S^\text{meas}  = \ln \tr(  \boldsymbol{\Pi}\z ^{t/\Delta t}\boldsymbol{\varrho}_\text{st}) + \delta S^\text{sta}.
\end{align}
The term $\delta S^\text{sta}$ accounting for the statistical error can be derived via the law of large numbers and is discussed in detail in the Supplementary Section~I.C.

\begin{figure}
	\includegraphics[width=8.6cm]{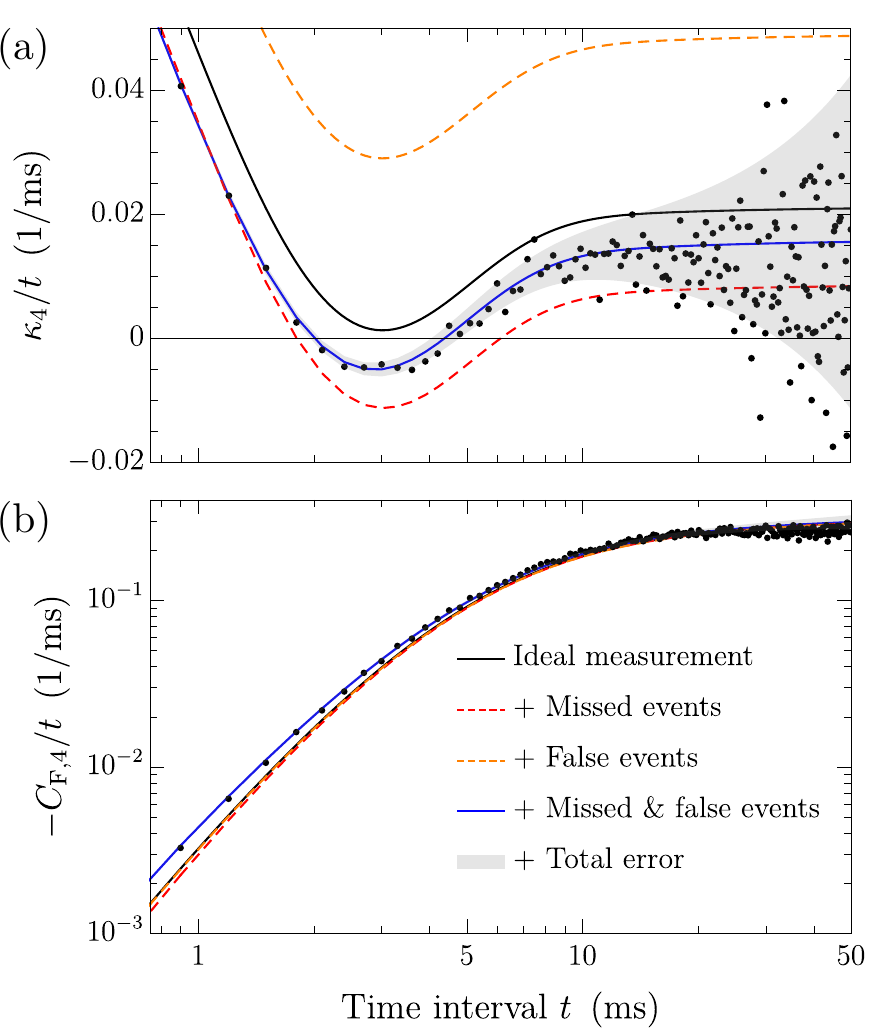}
	\caption{(a) Fourth ordinary cumulant $\kappa_4/t$ and (b) fourth factorial cumulant $-C_{\text{F},4}/t$ as a function of time $t$. Experimental data (black dots) is compared with theoretical calculations including: no error (black line), only a limited time resolution (red dashed line), only noise (orange dashed line), and both together (blue solid line). In gray, we indicate the statistical error due to a finite measurement time. The time resolution is $\Delta t=300\,\mu\text{s}$, the false-count rate of the bright state is $\Gamma_0^\text{false}=0.038\,\text{kHz}$, and the duration of the measurement is $T=369\,\text{s}$. The electron-tunneling rates are $\Gamma_\text{in}=0.346\,\text{kHz}$ and $\Gamma_\text{out}=0.334\,\text{kHz}$.} 
	\label{fig:5}
\end{figure}

By employing our model, we are finally able to explain the experimental results. We illustrate the impact of the different error sources separately in Fig.~\ref{fig:5}(a) for the fourth ordinary cumulant $\kappa_4$.
To obtain experimental data (black dots) with both a bad time resolution and many noise-induced false counts, we randomly deleted $95\%$ of all detected photons. The theoretical results (lines) are derived from Eq.~\eqref{eq:9new}. If we do not consider any error in our model (black solid line), the theory clearly deviates from the experiment (black dots). If we include only the noise-induced error (orange dashed line) then we overshoot, and if we include only the error due to the limited time resolution (red dashed line) then we undershoot. Only by considering both errors simultaneously (blue solid line), we find a nice agreement between theory and experiment. The continuous error bars that we obtained from $\delta S^\text{sta}$ [shaded area in  Fig.~\ref{fig:5}(a)] capture the statistical fluctuations around the blue curve due to the limited amount of data.
In contrast, for the fourth factorial cumulant $C_{\text{F},4}$ illustrated in Fig.~\ref{fig:5}(b), both the false and missed events have almost no effect, even though we used a poor time resolution $\Delta t =300\, \mu s$ and randomly deleted $95\%$ of all detected photons. In addition, a limited amount of data leads to only relatively weak statistical fluctuations, which we explain in more detail in the Supplementary Section~III.B.3.

To elucidate why factorial cumulants $C_{\text{F},m}$ possess a built-in ruggedness against measurement imperfections, we study the limit of small errors by performing a consistent perturbation expansion in the time resolution~$\Delta t$ and the false-count rates $\Gamma_{0}^\text{false}:=p(1\vert0)/\Delta t$ and $\Gamma_{1}^\text{false}:=p(0\vert1)/\Delta t$. Starting with the expression given in Eq.~\eqref{eq:8new} which is valid for arbitrarily strong measurement imperfections, we  find
\begin{align}\label{eq:10new}
\dot{\rho}\z=\big({\cal W}\z+{\cal W}^\text{miss}\z +{\cal W}^\text{false}\z \big) {\rho}\z,
\end{align}
where, in addition, we performed a partial trace over the detector degrees of freedom, $\rho\z=\sum_b \rho\z^{(b)}$.
Thus, the errors of missing ${\cal W}^\text{miss}\z$ and false events  ${\cal W}^\text{false}\z$ enter as effective corrections to the actual quantum dynamics encoded in ${\cal W}\z$. 
In particular, we find ${\cal W}^\text{miss}\z=-(z{-}1)\Delta t ({\cal J}_\text{in}{\cal J}_\text{out} {+}{\cal J}_\text{out}{\cal J}_\text{in})/2$ which describes successive tunneling-in (${\cal J}_\text{in}{=}{\cal P}_1 {\cal W} {\cal P}_0$) and tunneling-out (${\cal J}_\text{out}{=}{\cal P}_0 {\cal W} {\cal P}_1$)  events too close to each other to be resolved by the detector (similar to A and C in Fig.~\ref{fig:1}). This leads to missing counts. 
The false events due to noise are described by the diagonal matrix ${\cal W}^\text{false}\z=(z{-}1)\text{diag}\big(\Gamma^\text{false}_0, \Gamma^\text{false}_1\big)$. With rate $\Gamma^\text{false}_a$, the telegraph signal suffers from spurious switches to neighboring values $b\neq a$ and back again to $a$ (similar to B and D in Fig.~\ref{fig:1}).
Accordingly, we find for the generating function
\begin{align}\label{eq:11new}
S^\text{meas} = \ln \tr(e^{{\cal W}\z t \,+\,{\cal W}^\text{miss}\z t+\,{\cal W}^\text{false}\z t} \rho_\text{st}) + \delta S^\text{sta}, 
\end{align}
where the errors of missing (${\cal W}^\text{miss}\z$) and false (${\cal W}^\text{false}\z$) events still enter in a complicated way.  However, the expression simplifies considerably in the limit of short time intervals~$t$. Then, the systematic error reads
\begin{equation}\label{eq:12new}
\delta S^\text{sys} = (z-1) \, \big( \Gamma^\text{false} -  \Gamma^\text{miss}\big)\,t,
\end{equation}
with the mean rates $\Gamma^\text{false/miss}{=}{\pm}\partial_z {\tr}( {\cal W}\z^\text{false/miss} \rho_\text{st})\vert_{z=0}$.
As a result, the corrections due to false and missing events turn out to be Poisson like with positive and negative prefactors, respectively. 
This is true even for arbitrary times $t$ if both ${\cal W}^\text{false}\z\propto \mathds{1}$ and ${\cal W}^\text{miss}\z\propto \mathds{1}$, i.e., if the false and missed events happen independently of the 	quantum state.
In our experimental setup, however, the bright-state intensity fluctuates much more than the dark-state signal [see Fig.~\ref{fig:2}(a)], and, therefore, the false-count rates are heavily state dependent, $\Gamma_0^\text{false}\gg \Gamma_1^\text{false}$. Nonetheless, we find that Eq.~\eqref{eq:12new} also holds for all times $t$ if the electron tunneling rates fulfill $\Gamma_\text{in}\approx\Gamma_\text{out}$~(Supplementary Section~III.B).

With $\delta S^\text{sys}$ given in Eq.~\eqref{eq:12new}, the systematic error of both the ordinary  $\delta \kappa_m^\text{sys}= \partial_\chi^m \delta S^\text{sys}(e^\chi)\vert_{\chi=0}$ and factorial cumulants $\delta C_{\text{F},m}^\text{sys}=\partial_z ^m \delta S^\text{sys}(z)\vert_{z=1}$ can be determined. 
While the error of ordinary cumulants persists for all orders $m$, it is identically zero for factorial cumulants, $\delta C_{\text{F},m}^\text{sys}=0$ for all orders $m>1$.
Since it is highly unlikely that the mean rates of false and missing counts are known exactly, the systematic error of ordinary cumulants $\delta \kappa_m^\text{sys}$ cannot be corrected. Therefore, in this Letter, we suggest  that factorial cumulants $C_{\text{F},m}$ should always be used instead of ordinary cumulants $\kappa_m$ when analyzing telegraph signals. Not only do they provide a superior way to characterize the measured probability distribution~\cite{kambly2011factorial}.
The most striking advantage is that they automatically cancel out systematic errors $\delta S^\text{sys}$, so that detailed knowledge of the specific value of the error is not required anymore. Thereby, factorial cumulants push the limits set by typical detection errors. 

In summary, we demonstrated how quantum dynamics detected in real time can be evaluated 
by statistical means that are insensitive to typical, unavoidable experimental errors. The evaluation scheme is based on factorial cumulants, which are not influenced by any spurious signals caused by uncorrelated Poisson processes.  
Nevertheless, factorial cumulants contain the same information about the studied quantum system as ordinary cumulants~\cite{beenakker_counting_2001,kambly2011factorial,stegmann2015detection,stegmann2016short,kleinherbers2018revealing}. Our work opens up a new perspective to gain precision in the analysis of existing and future experimental data~\cite{hume_2011,natterer2017,saskin_narrow_2019,zhou2011detecting,choi2012single,chung2012single,xin2019concepts,miyamachi_2012,burzuri2018spin,guerlin_progressive_2007,matsuo_2020,gupta_2021,muckel_2021,pekola_single_2013,efros_1997,gustavsson2006counting,flindt2009universal,komijani2013counting,Ott_single_2016}. 
For charge fluctuations in a self-assembled quantum dot, we demonstrated error reduction by orders of magnitude. 
We emphasize that our approach is purely passive, i.e., it leaves the studied quantum dynamics unchanged and thus allows for a high-precision analysis, so that e.g., hidden quantum states, internal quantum transitions, and particle interactions can be revealed.

\begin{acknowledgments}
We thank C. Flindt for useful discussions. This work was supported by the Deutsche Forschungsgemeinschaft (DFG, German Research Foundation) under Project-ID 278162697 -- SFB 1242. P.S. acknowledges support from the German National Academy of Sciences Leopoldina (Grant No.~LPDS 2019-10).
\end{acknowledgments}

%

\end{document}


\title{
\textmd{Supplementary Information for}\\
``Pushing the limits in real-time measurements of quantum dynamics''
}
\author{E.~Kleinherbers}
\email{eric.kleinherbers@uni-due.de}
\affiliation{Faculty of Physics and CENIDE, University of Duisburg-Essen, 47057 Duisburg, Germany}
\author{P.~Stegmann}
\email{psteg@mit.edu}
\affiliation{Department of Chemistry, Massachusetts Institute of Technology, Cambridge, Massachusetts 02139, USA}
\author{A.~Kurzmann}
\affiliation{2nd Institute of Physics, RWTH Aachen University, 52074 Aachen, Germany}
\author{M.~Geller}
\affiliation{Faculty of Physics and CENIDE, University of Duisburg-Essen, 47057 Duisburg, Germany}
\author{A.~Lorke}
\affiliation{Faculty of Physics and CENIDE, University of Duisburg-Essen, 47057 Duisburg, Germany}
\author{J.~K{\"o}nig}
\affiliation{Faculty of Physics and CENIDE, University of Duisburg-Essen, 47057 Duisburg, Germany}

\date{\today}

\maketitle

\onecolumngrid

{
  \hypersetup{linkcolor=black}
  \vspace{-1cm}
  \tableofcontents
}

\newpage
\section{MODEL}
By measuring the telegraph signal, we gain information about the full counting statistics of the quantum system. In particular, we can infer the probability $P_N\meas(t)$ that $N$ events have been counted in a time interval of length $t$. However, the measured probability distribution $P\meas_N$ is inevitably overlaid by errors [see Fig.~\ref{fig:s1}(a)]
\begin{align}\label{eq:s1}
P_N\meas=\sum_{N^\prime=0}^N P_{N{-}N^\prime}\qs \delta P_{N^\prime}\sys+\delta P_N\sta,
\end{align}
where $P_N\qs$ describes the true events of the quantum system, $\delta P_N\sys$ accounts for systematic errors (false and missing events) and $\delta P_N\sta$ describes the statistical error due to a finite measurement time. Note, that the contributions $\delta P_N\sta$ and $\delta P_N\sys$ are no proper probability distributions, especially because they can take negative values. 
Whereas the systematic corrections are normalized to one $\sum_N \delta P_N\sys=1$, the stochastic corrections sum up to zero $\sum_N \delta P_N\sta=0$. 
To fully characterize the measured probability distribution $P_N\meas$, we employ \textit{ordinary cumulants} $\kappa_m\meas$ as well as \textit{factorial cumulants} $C_{\text{F},m}\meas$ of order $m\in\mathbb{N}$. They are conveniently derived from the generating function 
\begin{align}
\Sc\meas(z)=\ln \left(\sum_N z^N P_N\meas \right)=S\qs(z)+\delta S\sys(z) +\delta S\sta(z),
\end{align}
via $\kappa_m\meas= \partial_\chi^m S(e^\chi)\vert_{\chi=0}$ and $C_{\text{F},m}\meas=\partial_z^m S(z)\vert_{z=1}$. 
Here, we defined $\Sc\qs=\ln \left(\sum_N z^N P_N\qs \right)$, $\delta\Sc\sys=\ln \left(\sum_N z^N \delta P_N\sys \right)$, and $\delta S\sta=S\meas{-}S{-}\delta S\sys$. Both ordinary and factorial cumulants contain on their own the full information about the distribution $P_N\meas$. Nonetheless, higher-order factorial cumulants $C_{\text{F},m}$ turn out to be insensitive to measurement errors included in $\delta S\sys$ and, as a consequence, they are better suited to reveal the true quantum dynamics encoded in $S\qs$. 

For later use, we find it convenient to introduce the generating function $S\sys=S\qs+\delta S\sys$, which includes contributions from the quantum system plus systematic errors.
The corresponding probability distribution $P_{N}\sys= \sum_{N^\prime=0}^N P_{N{-}N^\prime}\qs \delta P_{N^\prime}\sys$ is, then, normalized and non-negative as required for a proper distribution.

In the following sections, we separately discuss the various contributions to the full counting statistics. In Sec.~\ref{section:sys}, we discuss the contribution $S\qs$ originating from the quantum system. Next, in Sec.~\ref{section:det}, we discuss how to model the systematic errors $\delta S\sys$ due to a noisy signal and a limited time resolution and, finally, in Sec.~\ref{section:stoch}, we discuss how to model the statistical error $\delta S\sta$ due to a finite measurement time $T$. 

\subsection{Quantum master equation} 
\label{section:sys}
In a counting experiment, the quantum system is continuously monitored with respect to a certain observable and the distinct measurement outcomes labelled with $a \in \lbrace1,2,..., M\rbrace$ are recorded in form of a telegraph signal as a function of time $t$ [see black line in Fig.~\ref{fig:s1}(a)]. 
The quantum system is described by a density matrix $\rho$ and the dynamics is governed by a generalized master equation $\dot\rho={\cal W} \rho$, where the so-called Liouvillian ${\cal W}$ is a $D$-dimensional superoperator acting on the supervector $\rho$.
Since the measurement is assumed to be projective, the density matrix can always be decomposed as $\rho=\sum_a \rho^a$ with $\rho^a={\cal P}_a \rho$ such that $p(a)=\tr \rho^a$ gives the probability that the measurement outcome is $a$. 
Here, ${\cal P}_a$ is a superoperator that projects the density matrix onto the subspace with measurement outcome $a$. It fulfills the orthogonality relation ${\cal P}_a {\cal P}_{a^\prime}=\delta_{aa^\prime} {\cal P}_a$ and also the completeness relation $\sum_a {\cal P}_a =\mathds{1}$ for the relevant subspace we are interested in.   Hence, coherences of states with different measurement outcomes are assumed to immediately decohere.
To obtain the full counting statistics, an additional counting degree of freedom $N\in \lbrace 0,1,2,...\rbrace$ has to be introduced in the density matrix $\rho_N(t)$ such that $P_N\qs(t)=\tr \rho_N(t)$ gives the probability that $N$ events have been counted until time $t$. 
The time evolution for the $N$-resolved density matrix $\rho_N(t)$ is, then, governed by the generalized master equation
\begin{align}\label{eq:mastereq_N}
\dot{\rho}_N(t)=\sum_{N^\prime}\sum_{a,a^\prime}\delta_{{N{-}N^\prime},n_{aa^\prime}} {\cal P}_a{\cal W}{\cal P}_{a^\prime} \rho_{N^\prime}(t).
\end{align}
Here, ${\cal P}_a{\cal W}{\cal P}_{a^\prime}$ projects out only those transitions that change the measurement outcome from $a^\prime$ to $a$, and $n_{aa^\prime}$ denotes the accompanied increase of the counter $N$. The Kronecker delta $\delta_{{N{-}N^\prime},n_{aa^\prime}}$, then, picks out only those values of $N^\prime$ that fulfill $N=N^\prime+n_{aa^\prime}$. The maximum increase of the counter $N$ per transition will be denoted as $n_\text{max}=\max\limits_{a,a^\prime} n_{aa^\prime}$. The initial condition at $t=0$ is given by  $\rho_N(0)=\delta_{N,0} \rho_\text{st}$, i.e. we assume the counter is initially set to $N=0$ and the system has already reached a stationary state $\rho_\text{st}$ with ${\cal W}\rho_\text{st}=0$. 
At this point, it is very beneficial to perform a $z$-transform $\rho\z=\sum_N z^N \rho_N$ to arrive at 
\begin{align}\label{eq:mastereq_z}
\dot{\rho}\z(t)=\sum_{a,a^\prime}(\Cz)_{aa^\prime} {\cal P}_a{\cal W}{\cal P}_{a^\prime} \rho\z(t)={\cal Z} \lbrack  {\cal W} \rbrack \rho\z(t)
\end{align}
with initial condition $\rho_z(0)=\rho_\text{st}$. Now, the master equation is decoupled in the so-called \textit{counting variable} $z$. The generator ${\cal Z} \lbrack  {\cal W} \rbrack $ is obtained via the operation ${\cal Z} \lbrack  {\cal A} \rbrack =\sum_{a,a^\prime}(\Cz)_{aa^\prime}{\cal P}_a{\cal A}{\cal P}_{a^\prime}$ which merely adds a factor $(\Cz)_{aa^\prime}=z^{n_{aa^\prime}}$ for those elements of ${\cal A}$ that connect the subspaces of $a$ and $a^\prime$. The matrix $\Cz$ will be referred to as the counting matrix because it contains the information about which transitions increase the counter $N$ by how much. 
Note, that by setting $z=1$, we effectively trace out the counter $N$ and get $\rho_1=\sum_N \rho_N \equiv \rho$.
The solution of the master equation Eq.~\eqref{eq:mastereq_z} is readily given by $\rho\z(t)=e^{{\cal Z} \lbrack  {\cal W} \rbrack  t}\rho_\text{st}$. The cumulant generating function is obtained via $S(z)=\ln \tr  e^{{\cal Z} \lbrack  {\cal W} \rbrack  t}\rho_\text{st}$. Using the inverse $z$-transform, we can obtain the full counting statistics via $P_N\qs(t)=N!\partial_z^N\tr \rho\z(t)\vert_{z=0}$. 

\subsection{Systematic errors}
\label{section:det}
\begin{figure}
	\includegraphics[width=1. \linewidth]{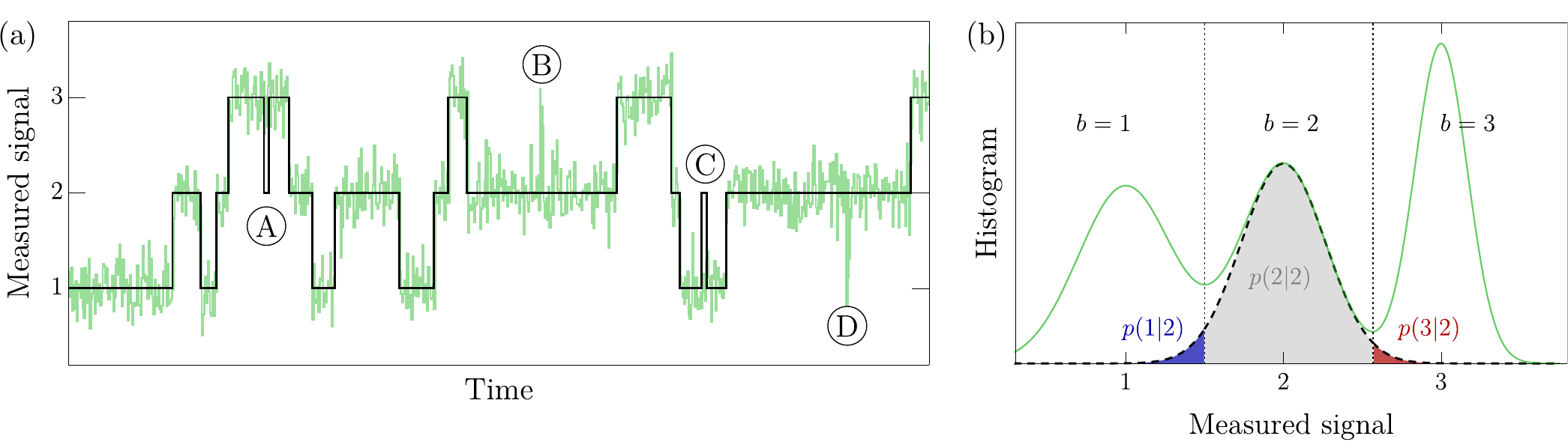}	
	\caption{Telegraph signal and histogram. (a)~The black line shows the true quantum dynamics of the system, where changes between the three measurement values $a\in\{1,2,3\}$ are recorded as a function of time. The green line shows a hypothetical measurement outcome.  Fast events (A and C) are missed due to the limited time resolution and additional false events (B and D) may occur due to noise. (b)~The histogram of the measured signal (green line) is broadened due to the noise. The two thresholds (dotted lines) define the three regions which will be associated with the measurement outcomes $b\in\{1,2,3\}$. The contribution of the true value $a=2$ is depicted as a dashed bell-shaped curve. Then, the overlap of the dashed curve with the respective region $b\in\{1,2,3\}$ (indicated blue, gray and red) characterizes the noise, i.e. the conditional probability $p(b\vert 2)$ that we measure $b$ given the true value is $a=2$.
	} 
	\label{fig:s1}
\end{figure}

A realistic counting experiment is inevitably subjected to systematic detection errors.
In Fig.~\ref{fig:s1}(a), a hypothetical measurement outcome (green) is depicted in comparison to the true quantum dynamics (black). We can identify two kinds of detection errors. Either some fast events are missed (A and C) due to a limited time resolution or some events are false (B and D) due to noise in the measured signal. 
Here, we propose a model that captures these errors at the level of the master equation, allowing an analytical estimation of their impact.  
Thus, we provide a valuable tool to verify whether theoretically predicted signatures in the counting statistics are actually measurable or not. The master equation reads
\begin{align}\label{eq:mastereq_error}
\rho\z^{a \cap b} (t+\Delta t)=p(b\vert a)\sum_{a^\prime,b^\prime} (\Cz)_{b b^\prime} {\cal P}_a e^{{\cal W}\Delta t} {\cal P}_{a^\prime} \rho\z^{a^\prime \cap b^\prime}(t),
\end{align}
where the initial stationary state is given by $\rho^{a\cap b}_\text{st}=p(b\vert a) {\cal P}_a \rho_\text{st}$.
Two major changes are built into Eq.~\eqref{eq:mastereq_error}:
\begin{itemize}
\item First, the measured signal is noisy and it may happen that the measured value $b\in\lbrace1,2,...,M\rbrace$ is false, i.e. it differs from the true value $a\in\lbrace1,2,...,M\rbrace$ of the quantum state. 
As a consequence, the dimensionality of the model increases by a factor $M$ because each quantum state can be associated with each measurement outcome $b$. In particular, the projected density matrix $\rho^a={\cal P}_a \rho$ for the true value $a$ must be artificially resolved into $\rho^a=\sum_b \rho^{a\cap b}$ such that $p(a\cap b)=\tr \rho^{a\cap b}$ describes the joint probability that the true value is $a$ and we measure $b$. Then, the noise-induced transitions can be fully described by the conditional probabilities $p(b\vert a)$ that we measure $b$ given the true value is $a$. Indeed, by performing the trace and setting $z=1$ in Eq.~\eqref{eq:mastereq_error}, we obtain for all times the well-known Kolmogorov definition of a conditional probability $p(b\vert a)=p(a\cap b)/p(a)$, where $p(a)=\tr \rho^a $ gives the probability that the true value is $a$~\footnote{Here, we used that i) the counting matrix is trivial $(\Cz)_{b b^\prime}=1$ when $z=1$, ii) that the projection operator ${\cal P}_{a^\prime} \rho^{a^\prime \cap b^\prime}=\rho^{a^\prime \cap b^\prime}$ is idempotent, iii) the original density matrix is retrieved via $\sum_{a^\prime,b^\prime}\rho^{a^\prime \cap b^\prime}=\rho$, and iv) the propagation of the density matrix by $\Delta t$ is given via $\rho^a(t+\Delta t)={\cal P}_a e^{{\cal W}\Delta t}\rho(t)$.}.
The transition probabilities $p(b\vert a)$ can be extracted directly from the measured telegraph signal. In Fig.~\ref{fig:s1}(b), we show for a concrete example with $a\in \{ 1,2,3 \}$ how $p(1\vert 2)$, $p(2\vert 2)$ and $p(3\vert 2)$ are obtained. After evaluating the histogram of the telegraph signal (green line), we have to identify the broadened peak (dashed bell-shaped curve) that originates from the true value $a=2$. Then, by choosing thresholds (dotted lines), we divide all possible outcomes into three sections that will be associated with the measured values $b\in \{ 1,2,3 \}$. Finally, the overlap of the dotted bell-shaped curve ($a=2$) with a given section $b\in \{ 1,2,3 \}$ determines the transition probabilities $p(1\vert 2)$, $p(2\vert 2)$ and $p(3\vert 2)$ (indicated in blue, gray, and red) up to a normalization factor that ensures $p(1\vert 2) {+} p(2\vert 2) {+} p(3\vert 2)=1$. 

\item Second, to account for missing events due to a limited time resolution $\Delta t$, we discretize the master equation in time steps of $\Delta t$. Then, the quantum degrees of freedom are stepwise propagated by ${\cal P}_a e^{{\cal W}\Delta t} {\cal P}_{a^\prime}$ which implicates a change of the true value from $a^\prime$ to $a$. The counting degree of freedom $N$, on the other hand, is updated each time step $\Delta t$ according to the counting matrix $(\Cz)_{b b^\prime}$. Here, however, only changes between the measured values from $b^\prime$ to $b$ --- not the true values from $a^\prime$ to $a$ --- influence the counter $N$. Most importantly, the counting matrix $(\Cz)_{b b^\prime}$ enters the coarse-grained Eq.~\eqref{eq:mastereq_error} only linearly.  Thereby, we guarantee that only the net change of the state after one integrated time step $\Delta t$ can influence the counter $N$. Multiple transitions happening on time scales smaller than $\Delta t$ are not individually counted. Thus, the maximum increase of the counter $N$ in one time step is limited by $n_{\text{max}}$.
Note, that the limited time resolution can cause, in general, not only missing but also false events. For example, the combination of several uncounted transitions can 
result in one effective transition that is, erroneously, counted by the detector. 
Nevertheless, to allow for a clear distinction between the miscounts induced by the detector noise and those due to the limited time resolution, we always refer to the latter as missing events.
\end{itemize}
Note, that the described modifications do not alter the underlying quantum dynamics. They merely describe the errors in the counting degree of freedom $N$. Hence, for $z=1$ Eq.~\eqref{eq:mastereq_error} reduces to the error-free equation $\rho(t+\Delta t)=e^{{\cal W} \Delta t} \rho(t)$. 

To solve Eq.~\eqref{eq:mastereq_error}, we rewrite it in an index-free notation. Therefore, we introduce $\rho^{(b)}=\sum_a \rho^{a\cap b}$ such that $\tr \rho^{(b)}$ describes the probability that the measurement outcome is $b$, independent of the true value $a$.
By rewriting the state as a large vector $\boldsymbol{\varrho}\z=\left( \rho\z^{(1)},...,\rho\z^{(b)},...,\rho\z^{(M)} \right)$ we obtain
\begin{align}\label{eq:mastereq_vector}
\boldsymbol{\varrho}\z (t+\Delta t)= {\cal F} \cdot \lbrack {\Cz} \otimes e^{{\cal W}\Delta t} \rbrack \boldsymbol{\varrho}\z(t),
\end{align}
where $\otimes$ denotes the tensor product between the $M$-dimensional counting matrix $\Cz=\sum_{b,b^\prime}  (\Cz)_{bb^\prime} \dyad{b}{b^\prime}$ in the basis of the detector states $\ket{b}$ and the $D$-dimensional propagator $e^{{\cal W}\Delta t}$.
Here, ${\cal F}=\sum_{a,b} p(b\vert a) \dyad{b}{b} \otimes {\cal P}_a$ is a diagonal matrix that associates the true value $a$ with the measured value $b$ with probability $p(b\vert a)$ and, thus, describes false events as well. The stationary state is given by $\boldsymbol{\varrho}_\text{st}=\left(\rho_\text{st}^{(1)},...,\rho_\text{st}^{(b)},...,\rho_\text{st}^{(M)}\right)$ with $\rho^{(b)}_\text{st}=\sum_a p(b\vert a) {\cal P}_a \rho_\text{st}$.
Now, the solution of Eq.~\eqref{eq:mastereq_vector} is given by $\boldsymbol{\varrho}\z(t)=\left({\cal F} \cdot \lbrack \Cz \otimes e^{{\cal W}\Delta t} \rbrack \right)^{t/\Delta t} \boldsymbol{\varrho}_\text{st}$, from which we obtain the generating function of the quantum system plus systematic errors via $S\sys(z)=\ln\tr \boldsymbol{\varrho}\z$, where the trace is defined by $\tr \boldsymbol{\varrho}\z=\sum_b \tr \rho\z^{(b)}$. 
The corresponding probability distribution is obtained via ${P}_N\sys(t)=N!\partial_z^N\tr \boldsymbol{\varrho}\z(t)\vert_{z=0}$. 
The correction to the generating function induced by the systematic errors is given by $\delta S\sys=S\sys -S$.

\subsubsection{Limited time resolution}
If the detector noise is negligible, the measured value $b$ always agrees with the true value $a$ and we can set $p(b\vert a)=\delta_{ab}$ to obtain a model for the error caused only by the limited time resolution. 
The condition $p(b\vert a)=\delta_{ab}$ implies that $\rho\z^{a\cap b} = \delta_{ab} \rho^{a\cap a}\z$ and thus $\rho\z=\sum_a \rho\z^{a\cap a}$. Then, by setting $b=a$ in Eq.~\eqref{eq:mastereq_error} and summing over $a$, we obtain after some algebra
\begin{align}\label{eq:mastereq_timeres}
\rho\z(t+\Delta t)=\sum_{a,a^\prime} (\Cz)_{a a^\prime} {\cal P}_a e^{{\cal W}\Delta t} {\cal P}_{a^\prime} \rho\z(t)={\cal Z}\lbrack e^{{\cal W}\Delta t} \rbrack \rho_z(t),
\end{align}
which has the same dimension as the original master equation Eq.~\eqref{eq:mastereq_z}. 
We emphasize that Eq.~\eqref{eq:mastereq_timeres} differs from simply integrating Eq.~\eqref{eq:mastereq_z} in time over the span $\Delta t$. 
The latter procedure would yield the solution $\rho\z(t+\Delta t)= e^{{\cal Z}\lbrack{\cal W}\rbrack \Delta t}  \rho_z(t)$, i.e., the counting operation would erroneously be applied at the level of the Liouvillian ${\cal Z}\lbrack{\cal W}\rbrack$.
It should, however, be applied at the coarse-grained level ${\cal Z}\lbrack e^{{\cal W}\Delta t} \rbrack$ instead. 
Therefore, the solution must be written as $\rho\z(t)={\cal Z}\lbrack e^{{\cal W}\Delta t} \rbrack^{{t}/{\Delta t}} \rho_\text{st}$ and the probability $P_N\sys(\Delta t)$ that $N$ events are counted in a time interval of length $\Delta t$ is given by $P_N\sys(\Delta t)=N! \partial_z^N \tr \lbrace {\cal Z}\lbrack e^{{\cal W}\Delta t} \rbrack \rho_\text{st} \rbrace \vert_{z=0}$. Since $\tr \lbrace {\cal Z}\lbrack e^{{\cal W}\Delta t} \rbrack \rho_\text{st} \rbrace $ is a polynomial of order $n_\text{max}$ in the counting variable $z$, we find that ${P}_N\sys(\Delta t)=0$ for $N>n_\text{max}$. Therefore, the model successfully describes that at most one event is detected during each time interval $\Delta t$ with a maximum increase $n_\text{max}$ of the counter $N$. 

\subsubsection{Noise}
In the limit of perfect time resolution $\Delta t \to 0$, we obtain from Eq.~\eqref{eq:mastereq_error} a simplified expression for the pure noise error. We assume that the noise probabilities take the form $p(b\vert a)=\Gamma_{ba} \Delta t {+} {\cal O}(\Delta t^2)$ for $b\neq a$, where $\Gamma_{ba}$ define the noise rates. Normalization demands $p(a\vert a)=1{-}\sum_{b\neq a} \Gamma_{b a} \Delta t {+} {\cal O}(\Delta t^2)$. 
Then, we obtain from Eq.~\eqref{eq:mastereq_error} for $b\neq a$ that $\rho\z^{a \cap b}(t)=0$ in the limit $\Delta t \rightarrow 0$. Hence, it is sufficient to determine the time evolution for $\rho\z^{a\cap a}(t)$.
By expanding $\rho\z^{a^\prime\cap b^\prime}(t)\approx  \Delta t \Gamma_{b^\prime a^\prime}  (\Cz)_{b^\prime a^\prime} \rho\z^{a^\prime \cap a^\prime}(t)$ for $b^\prime\neq a^\prime$ to the first non-vanishing order in $\Delta t$ and inserting it back into Eq.~\eqref{eq:mastereq_error} for $b=a$, we obtain after some rearrangement the following expression for the difference quotient
\begin{align}\label{eq:mastereq_noise}
\frac{\rho\z^{a \cap a}(t+\Delta t)- \rho\z^{a \cap a}(t)}{\Delta t}=  \sum_{a^\prime} (\Cz)_{aa^\prime} {\cal P}_a {\cal W} {\cal P}_{a^\prime}\rho\z^{{a^\prime} \cap {a^\prime}}(t)  + \sum_{b \neq a} \Gamma_{b a} \left[(\Cz)_{ab}(\Cz)_{ba}-1\right] \rho\z^{a \cap a}(t)+ {\cal O}(\Delta t).
\end{align}
Taking the limit $\Delta t \rightarrow 0$ and summing over $a$ we find for $\rho\z=\sum_a \rho\z^{a\cap a}$ the differential equation
\begin{align}
\dot{\rho}\z =\left({\cal Z}\lbrack {\cal W} \rbrack + {\cal N}\z \right)\rho\z,
\end{align}
where we define the noise matrix as ${\cal N}\z=\sum_{a,b\neq a} \Gamma_{ba}\left[(\Cz)_{ab}(\Cz)_{ba}{-}1\right]{\cal P}_a$.
Hence, the dimension is the same as in the original master equation Eq.~\eqref{eq:mastereq_z} and noise events appear as simple self loops with rates $\Gamma_{ba}$. Those self loops simulate noise-induced fluctuations $a\rightarrow b\rightarrow a$ from the true value $a$ to the false value $b$ and back again to $a$. In the process, the counter $N$ may increase due the changes $a\rightarrow b$ and/or $b\rightarrow a$ according to $(\Cz)_{ab}(\Cz)_{ba}$.

\subsubsection{Small errors}\label{sec:smallerrors}
If the detection errors are small, we find the following correction to the counting statistics 
\begin{align}\label{eq:mastereq_smallerrors}
\dot{\rho}\z (t)=\left( {\cal W}\z  + {\cal W}\z^\text{miss}+ {\cal W}\z^\text{false} \right)\rho\z(t),
\end{align}
where we use the notations ${\cal W}\z = {\cal Z}\lbrack{\cal W}\rbrack$ to denote the true dynamics, ${\cal W}\z^\text{miss}$ to describe the missing events in first order $\Delta t$ and $ {\cal W}\z^\text{false}$ to describe the false events in first order $\Gamma_{ba}$. 
To find the missing events ${\cal W}\z^\text{miss}$ in leading order, we transform Eq.~\eqref{eq:mastereq_timeres} into the interaction picture $\tilde{\rho}\z(t)=e^{- {\cal Z}\lbrack{\cal W}\rbrack t} \rho\z(t)$ with respect to the true counting statistics, i.e. we undo the counts originating from the true quantum dynamics. Then, we obtain for the difference quotient
\begin{align}
\frac{\tilde{\rho}\z(t+\Delta t)- \tilde{\rho}\z(t)}{\Delta t}&=e^{- {\cal Z}\lbrack{\cal W}\rbrack t} \left\{\frac{ e^{- {\cal Z}\lbrack{\cal W}\rbrack \Delta t} {\cal Z}\left[ e^{{\cal W} \Delta t} \right] -\mathds{1}}{\Delta t} \right\} e^{ {\cal Z}\lbrack{\cal W}\rbrack t} \tilde{\rho}\z(t)\\
&= e^{- {\cal Z}\lbrack{\cal W}\rbrack t} \left\{ -\frac{1}{2} \Delta t \left({\cal Z}\lbrack{\cal W}\rbrack^2 -{\cal Z}\lbrack{\cal W}^2\rbrack \right) + {\cal O}(\Delta t^2)\right\} e^{ {\cal Z}\lbrack{\cal W}\rbrack t}  \tilde{\rho}\z(t),
\end{align}
where we expanded the equation in $\Delta t$. By transforming back, we can identify ${\cal W}\z^\text{miss}=-\frac{\Delta t}{2} \left( {\cal Z}\lbrack{\cal W}\rbrack^2 {-}{\cal Z}\lbrack{\cal W}^2\rbrack \right)$ as a first-order correction to the true dynamics. 
Hence, the leading error ${\cal W}\z^\text{miss}$ originates from two fast subsequent events which are not counted properly, see A and C in Fig.~\ref{fig:s1}(a). 
The false events described by Eq.~\eqref{eq:mastereq_noise}, on the other hand, are already exact in first oder $\Gamma_{ba}$  such that ${\cal W}\z^\text{false}={\cal N}\z$.
Note, that for $z=1$ we have ${\cal W}_1^\text{miss}={\cal W}_1^\text{false}=0$ and ${\cal W}_1={\cal W}$ such that the quantum dynamics are unaltered. 

In the limit of small detection errors, one can  show that higher-order factorial cumulants $C_\text{F,m}\meas$ are generally less sensitive to errors than ordinary cumulants $\kappa_m\meas$. 
By solving Eq.~\eqref{eq:mastereq_smallerrors}, we find for the cumulant generating function ${S}\sys$ in first order $\Delta t$ and $\Gamma_{ba}$
\begin{align}
{S}\sys\approx \ln \tr e^{{\cal W}\z t} \rho_\text{st} + \frac{\mathrm{d}}{\mathrm{d} \eta} \frac{\tr e^{\left({\cal W}\z + \eta {\cal W}\z^\text{miss}+\eta {\cal W}\z^\text{false} \right)t} \rho_\text{st} }{\tr e^{{\cal W}\z t} \rho_\text{st} }\Bigg\vert_{\eta=0},
\end{align}
where we can identify via ${S}\sys=S\qs +\delta S\sys$ the contributions from the true quantum dynamics $S\qs=\ln \tr e^{{\cal W}\z t} \rho_\text{st}$ and the contributions due to detection errors $\delta S\sys$. In the short-time limit $\Gamma t \ll1$, where $\Gamma$ is a rate that represents the characteristic time scale of the system, $\delta S\sys$ reduces further to 
\begin{align}\label{eq:genfunc_error}
\delta S\sys = \sum_{n=0}^{\tilde{n}} \left(z^n-1\right) \left( \Gamma^{\text{false},n}-\Gamma^{\text{miss},n}\right) t,
\end{align}
where we defined the average rates $\Gamma^{\text{false},n}= \partial^n_z \tr ({\cal W}\z^\text{false} \rho_\text{st})/n!\vert_{z=0}$ and  $\Gamma^{\text{miss},n}=-\partial^n_z \tr ({\cal W}\z^\text{miss} \rho_\text{st})/n!\vert_{z=0}$ describing false counts and missing events that change the counter by $n$. The maximum error-induced change of the counter is limited by $\tilde{n}\leq 2n_\text{max}$. Interestingly, Eq.~\eqref{eq:genfunc_error} holds for all times~$t$ if the detection errors are independent of the quantum state, i.e. ${\cal W}\z^\text{miss}\propto \mathds{1}$ and ${\cal W}\z^\text{false}\propto \mathds{1}$. 
In particular, we find that the systematic error of factorial cumulants $\delta {C}_{\text{F},m}\sys=\partial_z^m \delta{S}\sys(z)\vert_{z=1}=0$ is zero for $m>\tilde{n}$ because Eq.~\eqref{eq:genfunc_error} is a finite polynomial in the counting variable $z$ of order $\tilde{n}$.  For ordinary cumulants, on the other hand, the systematic error $\delta\kappa\sys_m=\partial_\chi^m \delta{S}\sys(e^\chi)\vert_{\chi=0}$ remains nonzero for all orders $m$ because the generating function $\delta S\sys(e^\chi)$ contains powers of arbitrary order in $\chi$. 
In other words, the systematic errors of Eq.~\eqref{eq:genfunc_error} take the form of Poisson-like corrections (for $\tilde{n}=1$ it is in fact a stochastically independent Poisson distribution) to the true counting statistics, to which higher-order factorial
cumulants are inherently resistant to.

\subsection{Statistical error}
\label{section:stoch}
So far, we have described only systematic measurement errors included in $P_N\sys=\sum_{N^\prime=0}^N P_{N{-}N^\prime}\qs \delta P_{N^\prime}\sys$. Now, we propose a way to estimate also the statistical error $\delta P_N\sta=P_N\meas-P_N\sys$ due to a finite measurement time $T$.
\subsubsection{Finite measurement time}
To estimate the error due to a finite measurement time $T$, we reexamine how the measured probability $P_N\meas(t)$ is obtained from the telegraph signal. First, the telegraph signal of length $T$ gets divided into $K=T/t$ windows of equal length $t$. Then, we count the number of events in each window $N_k$ for $k\in \{ 1,...,K \}$ and perform a sample average $\text{S}\lbrack X \rbrack = \sum_{k=1}^K X_k/K$ over the quantity $X_k=\delta_{N N_k}$. 
The sample average $\text{S}\lbrack X \rbrack=P_N\meas$ can be identified with the measured probability that $N$ events are counted in a time interval of length $t$. 
In contrast, the expectation value of $X=\delta_{NN^\prime}$ for a random number $N^\prime$ can be evaluated as $\text{E}\lbrack X \rbrack =\sum_{N^\prime=0}^\infty  \delta_{N N^\prime}P_{N^\prime}\sys=P_N\sys$ and gives the systematic probability $P_N\sys$ that $N$ events will be counted in a time interval $t$. 
From the law of large numbers we know that the sample average $\text{S}\lbrack X \rbrack$ approaches the expected value $\text{E}\lbrack X \rbrack $ according to 
$ (\text{S}\lbrack X \rbrack-\text{E}\lbrack X \rbrack )^2 \sim \text{Var}\lbrack X \rbrack /K$
as the sample size $K$ increases, where the variance is given by
$\text{Var}\lbrack X \rbrack =\text{E}\lbrack X^2 \rbrack -\text{E}\lbrack X \rbrack ^2={P}_N\sys(1-{P}_N\sys).$
Hence, the statistical error is of the order of
\begin{align}\label{eq:pn_staerror}
 \delta P_N\sta=P_N\meas {-} {P}_N\sys  \sim \sqrt{\frac{t}{T} {P}_N\sys\left(1-{P}_N\sys\right)}.
 \end{align} 
To model the finite measurement time, we assume that the measured probability distribution $P_N\meas$  always includes random fluctuations $\eta_N$ superimposed on the systematic probability distribution ${P}_N\sys$. We obtain $P_N\meas=\left({P_N\sys+\eta_N}\right)/{\sum_{N^\prime}\left( P_{N^\prime}\sys + \eta_{N^\prime}\right)}$, where we ensured normalization.
Thus, the statistical error reads
\begin{align}\label{eq:pn_staerror_random}
\delta P_N\sta= {\eta_N- {P}_{N}\sys \sum_{N^\prime}\eta_{N^\prime}},
\end{align}
in leading order in the small fluctuations $\eta_N$. 
The random fluctuations $\eta_N$ with $N=0,1,...$ are described by independent stochastic variables $\ev{\eta_N \eta_{N^\prime}}_\eta \propto \delta_{N N^\prime}$ drawn from Gaussian probability distributions with zero means $\ev{\eta_N}_\eta=0$ and finite variances $\sigma_N^2:=\ev{\eta_N^2}_\eta=\frac{t}{T} {P}_N\sys(1-{P}_N\sys) $ inspired by  Eq.~\eqref{eq:pn_staerror}.
The brackets $\ev{...}_\eta=\prod\limits_N \int \frac{\mathrm{d}\eta_N}{\sqrt{2\pi}\sigma_N} ... ~e^{-\eta_N^2/(2\sigma_N^2)}$ indicate the average over all possible fluctuations. 

Now, the statistical error of  any observable  $A\meas$ can be approximated as
\begin{align}\label{eq:obs_staerror}
\delta A\sta=A\meas-A\sys\approx \sum_N \frac{\partial A\sys}{\partial P_N\sys}\delta P\sta_N,
\end{align}
in the leading order in $\eta_N$. Here, the observable $A\sys=A\sys(P_0\sys, P_1\sys,...)$ is expressed as a function of the probabilities $P_N\sys$.
The mean squared deviation $(\Delta A\sta)^2=\ev{(\delta A\sta)^2}_\eta$ of those random fluctuations describes a typical statistical error of the measurement outcome and can be evaluated as
\begin{align}\label{eq:obs_msd}
\left(\Delta A\sta \right)^2&=\frac{t}{T}\sum_N {P}_N\sys(1-{P}_N\sys) \left[ \sum_{N^\prime} \frac{\partial A\sys}{\partial P_{N^\prime}\sys}\left(\delta_{N N^\prime} -P_{N^\prime}\sys \right)\right]^2,
\end{align}
where we used Eq.~\eqref{eq:pn_staerror_random}-\eqref{eq:obs_staerror} and took into account that $\ev{\eta_N \eta_{N^\prime}}_\eta =\delta_{NN^\prime}\sigma_N^2$.

Alternatively, we can express the observable $A\sys=A\sys(Y_1\sys,Y_2\sys,...)$ as a function of any other set of variables $Y_k\sys$ such that the statistical error becomes  $\delta A\sta\approx \sum_k  \frac{\partial A\sys}{\partial Y_k\sys} \delta Y_k\sta $. Then, the mean squared deviation is obtained via
\begin{align}\label{eq:obs_msd_any}
\left(\Delta A\sta \right)^2&=\sum_{k, k^\prime} \frac{\partial A\sys}{\partial Y\sys_k} \frac{\partial A\sys}{\partial Y\sys_{k^\prime}}\ev{\delta Y_k\sta \delta Y_{k^\prime}\sta}_\eta,
\end{align}
where $\ev{\delta Y_k\sta \delta Y_{k^\prime}\sta}_\eta$ denotes the covariance of the fluctuations $\delta Y_k\sta=\sum_N \frac{\partial Y\sys_k}{\partial P_N\sys}\delta P\sta_N$.

We are interested in the statistical error of ordinary $\kappa_m\meas$ and factorial cumulants $C_{\text{F},m}\meas$. 
For those observables it is convenient to express the errors in terms of ordinary moments $Y_k\sys=M_k\sys:=\sum_N N^k P_N\sys $ because the covariance $\ev{\delta M_k\sta \delta M_{k^\prime}\sta}_\eta\approx \frac{t}{T} \left( M_{k+k^\prime}\sys-M_k\sys M_{k^\prime}\sys\right)$ takes a particularly simple form. Here, we approximated $\sigma_N^2 \approx \frac{t}{T} P_N\sys$ since the statistical error is most significant for long time intervals, where $P_N\sys\ll1 $. We find
\begin{align}\label{eq:obs_msd_mom}
\left(\Delta A\sta \right)^2=\frac{t}{T} \sum_{k,k^\prime} \frac{\partial A\sys}{\partial M_k\sys}\frac{\partial A\sys}{\partial M_{k^\prime}\sys}\left( M_{k+k^\prime}\sys-M_k\sys M_{k^\prime}\sys\right),
\end{align}
which has the great advantage over Eq.~\eqref{eq:obs_msd} that for the observables $A\meas\in \{\kappa_m\meas,C_{\text{F},m}\meas \} $ the sums each have only $m$ terms. 

\subsubsection{Errors in the long-time limit} 
In the long-time limit, the expression for the statistical error of ordinary and factorial cumulants can be significantly simplified. 
To estimate the error, we examine the cumulant generating function, which can be approximated by a Gaussian distribution  $S\sys(e^\chi)\approx \kappa_1\sys \chi +\frac{1}{2} \kappa_2\sys \chi^2$ in the long-time limit due to the central limit theorem. 
Hence, only the first two cumulants describing the mean value  $\kappa_1\sys$ and the variance $\kappa_2\sys$ have to be considered. 
By using $\delta S\sta=S\meas-S\sys$ we find with Eq.~\eqref{eq:pn_staerror_random}
\begin{align}
\delta S\sta (e^\chi)&=\sum_N \eta_N \left[ e^{N\chi-S\sys(e^\chi)} -1\right]
\approx \sum_N \eta_N \left[ e^{(N-\kappa_1\sys)\chi -\frac{1}{2} \kappa_2\sys \chi^2}-1 \right].
\end{align}
Therefore, we find for the statistical error of ordinary cumulants $\delta \kappa_m\sta=\partial_\chi^m \delta S\sta \vert_{\chi=0}$ the following covariance 
\begin{align}
\ev{\delta \kappa_m\sta \delta \kappa_{m^\prime}\sta}_\eta=\delta_{m m^\prime} \frac{t}{T} m! (\kappa_2\sys)^m,
\end{align}
where, again, we used the approximation $\sigma_N^2 \approx \frac{t}{T} P_N\sys$. 
Hence, the statistical errors of all ordinary cumulants are independent of each other since $\ev{\delta \kappa_m\sta \delta \kappa_{m^\prime}\sta}_\eta\propto\delta_{m m^\prime}$. By employing the linear relation between factorial and ordinary cumulants $C_{\text{F},m}\sys=\sum_{k=1}^m s_1(m,k)\kappa_k\sys$ with $s_1(m,k)$ being the Stirling numbers of the first kind, we can estimate the error of factorial cumulants as $\delta C_{\text{F},m}\sta=\sum_{k=1}^m s_1(m,k)\delta\kappa_k\sta$. In leading order in $t$, we find 
\begin{align}\label{eq:cum_msd}
\Delta C_{\text{F},m}\sta&=  \Delta \kappa_{m}\sta = \sqrt{\frac{m! t }{T}} ({\kappa}_2\sys)^\frac{m}{2} \propto t^{\frac{m+1}{2}},
\end{align}
where we used $s_1(m,m)=1$.
Apparently, the absolute errors of factorial and ordinary cumulants become equal $\Delta C_{\text{F},m}\sta =  \Delta \kappa_{m}\sta $ in the limit of long times $t$. Furthermore, the relative error of the cumulants $\Delta C_{\text{F},m}\sta/{C}_{\text{F},m}\sys \propto  \Delta \kappa_{m}\sta/{\kappa}_m\sys \propto t^{(m-1)/2}$ increases with time according to a power law for $m>1$. 

\section{OPTICAL READOUT}

\begin{figure}
        \includegraphics[width=1. \linewidth]{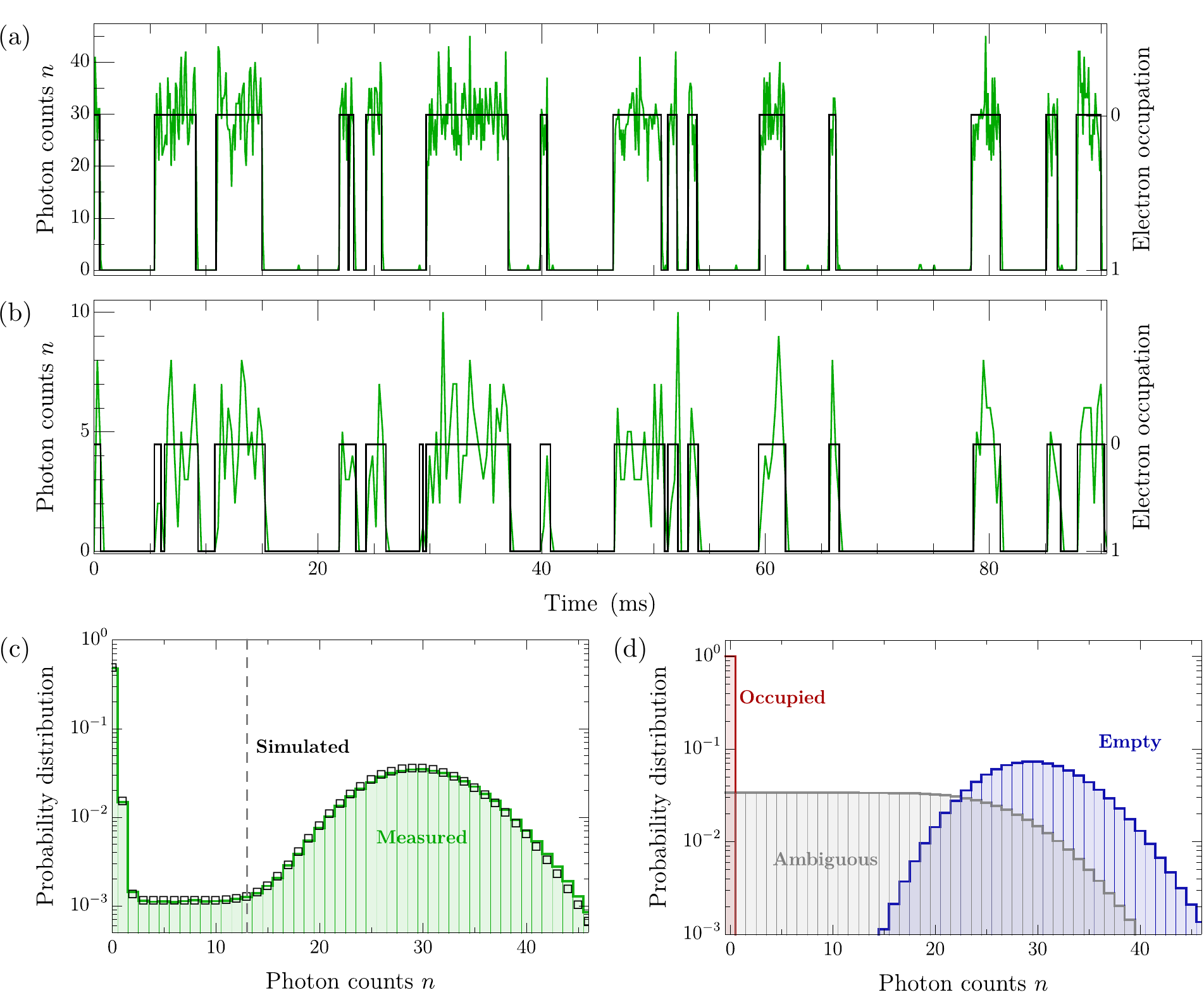}	
		\caption{Telegraph signal and histogram. (a)-(b)~The green line shows the measured number of photons as a function of time with a binning time of (a) $\Delta t=100\,\mu\text{s}$ and (b) $\Delta t=300\,\mu\text{s}$. By using a threshold of (a) $n_\text{th}=13$ and (b) $n_\text{th}=0$, the binned photon stream is transformed into a binary signal describing the electron occupation of the quantum dot (black line). In (b), $95~\%$ of all photons have been randomly deleted. (c)~The histogram (green) of the binned photon stream from (a) is depicted in a logarithmic scale. The broad peak at $n\approx 30$ indicates that the quantum dot is fluorescent (empty), while the narrow peak at $n=0$ indicates that the quantum dot emits no  photons (occupied). The dashed line indicates the threshold $n_\text{th}=13$. The photon statistics  (open black squares) simulated by means of the master equation Eq.~\eqref{eq:mastereq_twostates} agrees nicely with the measured photon statistics (green histogram). The parameters for the simulation are $\Gamma_\text{X}=298\,\text{kHz},\Gamma_\text{bg}=0.288\,\text{kHz}, \Gamma_\text{in}=0.346\,\text{kHz}$ and $\Gamma_\text{out}=0.334\,\text{kHz}$. (d)~The photon distribution can be decomposed into three parts (red, blue and gray), cf. Eq.~\eqref{eq:pn_decomp}, where each part describes a normalized probability distribution. The distribution $p_n^\text{occ}$ (red) captures the absence of photon emission if the quantum dot is occupied.  The Poisson distribution $p_n^\text{emp}$ (blue) characterizes the photon emission if the quantum dot is empty. If the electron occupation changes within a time interval $\Delta t$, an assignment becomes ambiguous. Then, the photon emission is characterized by a plateau-shaped distribution $p_n^\text{amb}$ (gray).} 
	\label{fig:s2}
\end{figure}

\subsection{From photon to electron statistics}
The self-assembled quantum dot is coupled to a charge reservoir via a tunneling barrier and can dynamically change its occupation between \textit{empty} and $\textit{singly occupied}$. Whenever the quantum dot is empty, an excitonic transition is resonantly driven by a laser. The single photons emitted from the quantum dot are detected in a time-resolved manner by an avalanche photo diode (resolution of $350~\text{ps}$).  These individually detected photons are recorded as a photon stream by a time-to-digital converter having a minimal resolution (jitter) of $81~\text{ps}$. 
By choosing a specific binning time $\Delta t$, that means counting all photons in a certain time bin of length $\Delta t$, we obtain a telegraph signal as shown in Fig.~\ref{fig:s2}(a)-(b)(green line). Thus, the chosen binning time is equal to the time resolution $\Delta t$ of the measurement. 
While in Fig.~\ref{fig:s2}(a) the time resolution is $\Delta t=100\,\mu\text{s}$, we choose a larger binning time $\Delta t=300\,\mu\text{s}$ in Fig.~\ref{fig:s2}(b) since $95\%$ of all photons have been randomly deleted (cf. Fig.~{\color[rgb]{1,0,0}4} of the main text). The corresponding probability distribution $p_n=p_n(\Delta t)$ that $n$ photons are detected in a binning time interval $\Delta t$ is depicted in Fig.~\ref{fig:s2}(c) which is evaluated for the photon stream of Fig.~\ref{fig:s2}(a). There, the narrow peak at $n=0$ indicates the dark state (occupied dot) and the broad peak indicates the bright, fluorescent state (empty dot). Next, we need a threshold $n_\text{th}$ to transform the photon signal [green line in Fig.~\ref{fig:s2}(a)] into a binary signal of the electron occupation [black line in Fig.~\ref{fig:s2}(a)]. The quantum dot is occupied for $n \le n_\text{th}$ and empty for $n> n_\text{th}$. In the following, we present a unique way to extract this threshold from the photon statistics. 
The measured photon distribution can be written into the form $p_n=\sum_{k=0}^n p_{n-k}^\text{qd} p_k^\text{bg}$. The small background of photons is characterized by a stochastically independent Poisson distribution $p_k^\text{bg}=(\Gamma_\text{bg} \Delta t)^k/k! e^{-\Gamma_\text{bg} \Delta t}$ with a constant rate $\Gamma_\text{bg}\approx288\,\text{Hz}$. The majority of the measured photons is emitted from the quantum dot. We find for the corresponding probability distribution the decomposition
\begin{align}\label{eq:pn_decomp}
p_n^\text{qd}=\alpha p_n^\text{occ} + \beta p_n^\text{emp} +(1-\alpha-\beta)p_n^\text{amb} ,
\end{align}
with coefficients $\alpha$ and $\beta$ fulfilling $0<\alpha,\beta<1$. Here, $p_n^\text{occ}$ and $p_n^\text{emp}$ are two peaked distributions and $p_n^\text{amb}$ describes a plateau, see Fig.~\ref{fig:s2}(d). If the quantum dot remains the whole time $\Delta t$ in the dark state (occupied dot), no photons are emitted and $p_n^\text{occ}=\delta_{n,0}$ (red histogram). If it remains in the bright state (empty dot), the emitted photons turn out to be Poisson distributed and $p_n^\text{emp}=(\Gamma_\text{X}\Delta t)^n/n! e^{-\Gamma_\text{X}\Delta t}$ with $\Gamma_\text{X}\approx298\,\text{kHz}$ (blue histogram).  However, sometimes it may happen that the quantum-dot occupation changes within one time interval $\Delta t$. Then, photons are emitted only for a certain fraction $\varepsilon$ of the time interval with $0{<}\varepsilon{<}1$. The probability distribution $p_n^\text{amb}$ (gray histogram) for those ambiguous cases can be estimated as
\begin{align}
p_n^\text{amb}=\int_0^{1} \mathrm{d}\varepsilon \frac{( \Gamma_\text{X} \varepsilon \Delta t)^n}{n!}e^{-\Gamma_\text{X} \varepsilon  \Delta t}=\frac{n!-\Gamma(1+n,\Gamma_\text{X} \Delta t)}{ n! ~\Gamma_\text{X} \Delta t},
\end{align}
where $\Gamma(a,z)=\int_z^\infty \mathrm{d}t\, t^{a-1}e^{-t} $ is the upper incomplete gamma function.  Here, we assumed that all possible fractions $\varepsilon$ are equally probable. The resulting distribution has a characteristic plateau for $n\ll \Gamma_\text{X} \Delta t$ with  $p_n^\text{amb}=(1{-}e^{-\Gamma_\text{X} \Delta t})/(\Gamma_\text{X} \Delta t)$. By choosing the threshold $n_\text{th}$ such that $\sum_{n=0}^{n_\text{th}}p_n^\text{amb}\approx 1/2$, we ensure that the ambiguous cases are evenly distributed among the possible outcomes empty and occupied. In this way, we can systematically extract a threshold $n_\text{th}$ from the photon statistics and obtain a telegraph signal of the quantum dot occupation.

\subsection{Characterizing measurement errors}
Now, we can characterize the errors of the optical readout.
The great benefit of an optical readout is its fine and tunable binning time $\Delta t$. The noise, on the other hand, is characterized by the transition probabilities $p(b\vert a)$ for noise-induced events, i.e. the conditional probability that we measure $b$ given the true value is $a$. 
In an optical readout, the probabilities  $p(b\vert a)$ strongly depend on $\Delta t$ and can be extracted via 
\begin{align}\label{eq:noiseprob}
p(0\vert 1)&=\sum_{n>n_\text{th}}\left( \sum_{k=0}^{n} p_{n-k}^\text{occ}p_{k}^\text{bg}\right)=1-\frac{\Gamma(1+n_\text{th},\Gamma_\text{bg} \Delta t)}{n_\text{th}!}, \\
p(1\vert 0)&=\sum_{n\le n_\text{th}}\left( \sum_{k=0}^{n} p_{n-k}^\text{emp}p_{k}^\text{bg}\right)=\frac{\Gamma(1+n_\text{th},\Gamma_\text{X} \Delta t+\Gamma_\text{bg} \Delta t)}{n_\text{th}!}.
\end{align} 
Noise-induced events only happen when the number of photons is below (above) the threshold $n_\text{th}$ although the quantum dot is in fact empty (occupied). The noise from the photon background is small, see Fig.~\ref{fig:s2}(a). Therefore, the probability $p(0\vert 1)$ to falsely assign $b=0$ (empty) although we have $a=1$ (occupied) can, in most cases, be neglected. Interestingly, the plateau probability distribution $p_n^\text{amb}$ does not influence the noise probabilities, but it is rather a manifestation of the error due to a limited time resolution $\Delta t$.
Finally, the statistical error due to a finite measurement time is fully characterized by the length of the telegraph signal which is given by $T=369\,\text{s}$.

\subsection{Modelling photon statistics}
\begin{figure}
	\includegraphics[width=1. \linewidth]{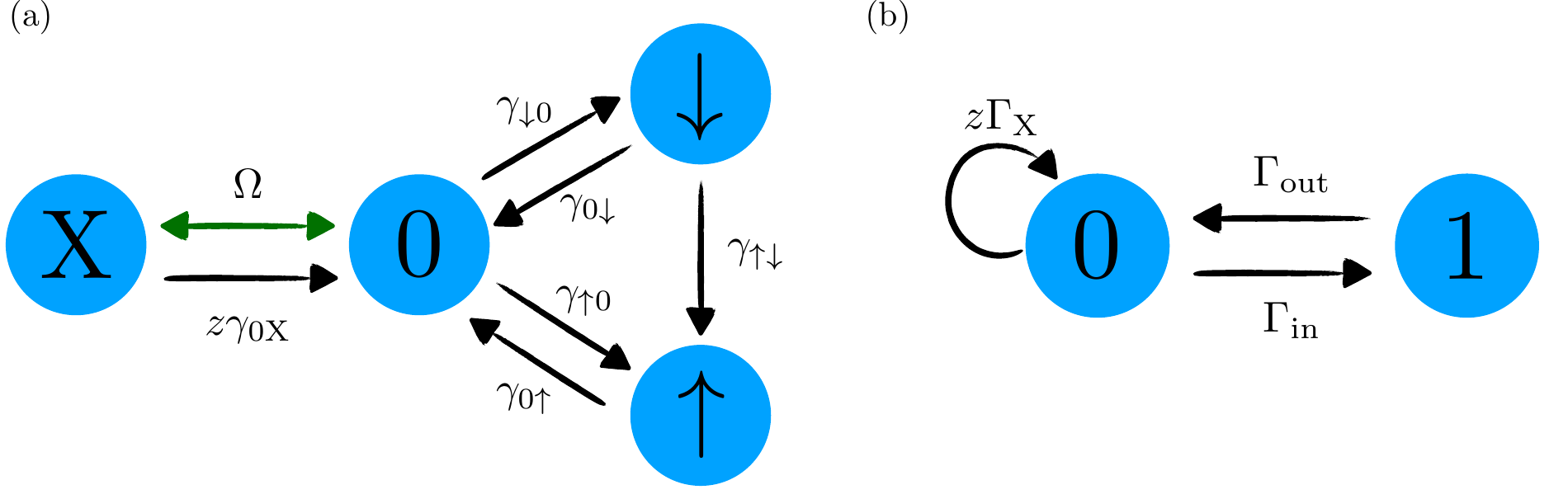}	
        \caption{Modelling photon statistics. (a)~The full dynamics of the quantum dot system is shown. Here, electrons with spin $\sigma \in \{\downarrow,\uparrow\}$ can tunnel into and out of the quantum dot with rates $\gamma_{\sigma 0}$ and  $\gamma_{0\sigma}$. Spin relaxation from the spin-down state $\ket{\downarrow}$  to the spin-up state $\ket{\uparrow}$ is described by the rate $\gamma_{\uparrow \downarrow}$. In addition, the exciton state $\ket{\text{X}}$ can be coherently driven by a laser with the Rabi frequency $\Omega$ whenever the quantum dot is empty $\ket{0}$. A relaxation from the excited state with rate $\gamma_{0\text{X}}$ results into the emission of a photon which is counted by the counting variable~$z$.  (b)~The system effectively reduces to a two state system with rates $\Gamma_\text{in},\Gamma_\text{out}$ and $\Gamma_\text{X}$, if both the spin relaxation dominates the electron time scales $\gamma_{\uparrow \downarrow}\gg\gamma_{\sigma 0},\gamma_{0\sigma}$ and the photon-detection rate  is only a small fraction of the photon-emission rate $\Gamma_\text{X}\ll\gamma_{0\text{X}},\Omega$. } 
	\label{fig:s3}
\end{figure}
The full dynamics of the quantum-dot system is depicted in Fig.~\ref{fig:s3}(a). The self-assembled quantum dot can be empty $\ket{0}$ or occupied $\ket{\sigma}$ by a single electron with spin $\sigma\in\lbrace\downarrow,\uparrow\rbrace$. The tunneling rates  $\gamma_{\sigma0}$  and $\gamma_{0\sigma}$ describe an electron entering and leaving the quantum dot, respectively. An applied magnetic field $B$ lifts spin degeneracy such that $E_\downarrow>E_\uparrow$ since InAs posses a negative $g$ factor. The quantum-dot spin can relax from the spin-down state $\ket{\downarrow}$ to the spin-up state $\ket{\uparrow}$ with the spin-relaxation rate $\gamma_{\uparrow\downarrow}$, where the phonon-assisted spin-flip mechanism is mediated by spin-orbit coupling~\cite{khaetskii_2001}.
Finally, for an optical readout of the electron occupation, the exciton state $\ket{\text{X}}$ is resonantly driven by an infrared laser whenever the quantum dot is empty. 
\subsubsection{Why are the emitted photons Poisson distributed?}
To understand why the emitted photons $p_n^\text{emp}$ are Poisson distributed if the quantum dot is empty, we study the isolated subsystem spanned by the states $\ket{0}$ and $\ket{\text{X}}$. Then, the optical drive can nicely be described by an optical Bloch equation for the density matrix $\rho\z$ in the rotating frame. The equation takes the form of a Lindblad equation
\begin{align}\label{eq:mastereq_bloch}  
\dot{\rho}\z=\frac{1}{i \hbar} \left[ H,\rho\z \right] + \gamma_{0\text{X}} \left(z L_{0\text{X}}\rho\z L^\dagger_{0\text{X}}-\frac{1}{2}\lbrace L^\dagger_{0\text{X}} L_{0\text{X}}, \rho\z \rbrace\right)+\frac{\gamma_{\phi}}{2} \left(L_{\phi} \rho\z L^\dagger_{\phi}-\frac{1}{2}\lbrace L^\dagger_{\phi} L_{\phi}, \rho\z \rbrace\right),
\end{align} 
where $z$ denotes the photon counting variable. Here, the coherent optical excitation is described by the Hamiltonian $H=\Omega \sigma_x/2$ with Pauli matrices $\sigma_x,\sigma_y,\sigma_z$ in the basis $\lbrace \ket{0},\ket{\text{X}}\rbrace$. The Rabi frequency $\Omega$ can be tuned with the laser power. In addition, we phenomenologically include the relaxation of the exciton state with rate $\gamma_{0\text{X}}$ and Lindblad operator $L_{0\text{X}}=(\sigma_x+i \sigma_y)/2$ as well as pure dephasing with rate $\gamma_\phi$ and Lindblad operator $L_\phi=\sigma_z$. The former process results in 
the emission of a single photon, which increases the photon counter from $n$ to $n+1$. Hence, the respective transition is multiplied with the photon counting variable $z$. Note, that  $T_1=\gamma_{\text{0X}}^{-1}$ describes the exciton lifetime and $T_2=(\gamma_\phi+\gamma_{\text{0X}}/2)^{-1}$ the dephasing time. 
By solving Eq.~\eqref{eq:mastereq_bloch} for $\rho\z(t)$  with an initial stationary state (obtained by $\dot{\rho}_1=0$), we can calculate the probability distribution $p_n^\text{emp}(t)=n! \partial_z^n \tr \rho\z(t) \vert_{z=0}$  that $n$ photons have been emitted from the empty quantum dot in a time interval $t$. The specific form of $p_n^\text{emp}(t)$ is, in general, complicated. 
However, since the photon detector measures only a small fraction $\alpha \approx 10^{-3}$ of all emitted photons from the quantum dot (due to a small collection efficiency of the optical setup), we have to modify Eq.~\eqref{eq:mastereq_bloch} according to $z\rightarrow \ (1-\alpha) + \alpha z$. Then, we obtain in very good approximation the cumulant generating function $S^\text{emp}=\ln \tr \rho_z (t)\approx \Gamma_\text{X} t (z{-}1) +{\cal{O}}(\alpha^2)$ of a Poisson distribution $p_n^\text{emp}(t) =(\Gamma_\text{X} t)^n/n! e^{-\Gamma_\text{X} t}$.  The effective photon-detection rate $\Gamma_\text{X}$ is obtained via $\Gamma_\text{X}= {\alpha}{\Omega^2 T_2}/({2{+}2T_1T_2\Omega^2})$. Hence, due to the fact that the detector is blind to a large fraction of photons, the measured photon statistics is sufficiently described by a single rate $\Gamma_\text{X}$ and specific details of the coherent excitation are irrelevant for the description. 
\subsubsection{Effective two-state system}
It turns out that the photon statistics can be sufficiently described by an effective two-state system, where the quantum dot is either empty $ \ket{0}$ or occupied $\ket{1}$, see Fig.~\ref{fig:s3}(b). Since the optical transitions are on a much faster time scale than electron tunneling, $\Omega,\gamma_{0\text{X}}\gg \Gamma_\text{X}\gg \gamma_{0\sigma}, \gamma_{\sigma0}$, we can replace the optical part of the dynamics by a simple self loop with rate $z \Gamma_\text{X}$ that starts and ends at the state $\ket{0}$, see Fig.~\ref{fig:s3}(b). This simulates the effective Poisson process with rate $\Gamma_\text{X}$ which has been discussed above. Furthermore, here a strong magnetic field of $B=10\,\text{T}$ has been applied, such that $\gamma_{\uparrow\downarrow}\gg\gamma_{0\sigma} $~\cite{kurzmann2019optical}. Thus, there is a negligible probability to find the system in the spin-down state~$\ket{\downarrow}$ and we can identify $\ket{1}\equiv\ket{\uparrow}$. The effective tunneling rates are obtained via $\Gamma_\text{in}\approx s( \gamma_{\downarrow 0}+ \gamma_{\uparrow 0})$ and $\Gamma_\text{out}\approx \gamma_{0\uparrow}$. The factor $s=\left[1{+}(1{+}T_1 T_2\Omega^2)^{-1}\right]/2\le1$ takes optical blocking into account, because whenever the system is in the exciton state, no electron can tunnel in. We find for the time evolution of the density matrix $\rho\z=(\mel{0}{\rho\z}{0},\mel{1}{\rho\z}{1})$ the following master equation 
\begin{align}
\dot{\rho}\z =\begin{pmatrix}
-\Gamma_\text{in} + \Gamma_\text{X} (z{-}1) & \Gamma_\text{out}  \\
\Gamma_\text{in} &  -\Gamma_\text{out}  
\end{pmatrix} \rho\z.
\label{eq:mastereq_twostates}
\end{align}
Note, that coherences $\mel{0}{\rho}{1}=\mel{1}{\rho}{0}=0$ are prohibited by a super-selection rule~\cite{wick_1952}. The resulting distribution $p_n^\text{qd}(t)=n! \partial_z^n \tr \rho_z(t) \vert_{z=0}$ describes photon emission from the quantum dot.  If we use the two-state model from Eq.~\eqref{eq:mastereq_twostates} and add the photon background $p_n=\sum_{k=0}^n p_{n-k}^\text{qd} p_k^\text{bg}$, a nice agreement between theory (open black squares) and the measured data (green histogram) is established, see Fig.~\ref{fig:s2}(c).

\section{QUANTUM-DOT SYSTEM}
\subsection{Quantum dynamics}
\begin{figure}
        \includegraphics[width=1. \linewidth]{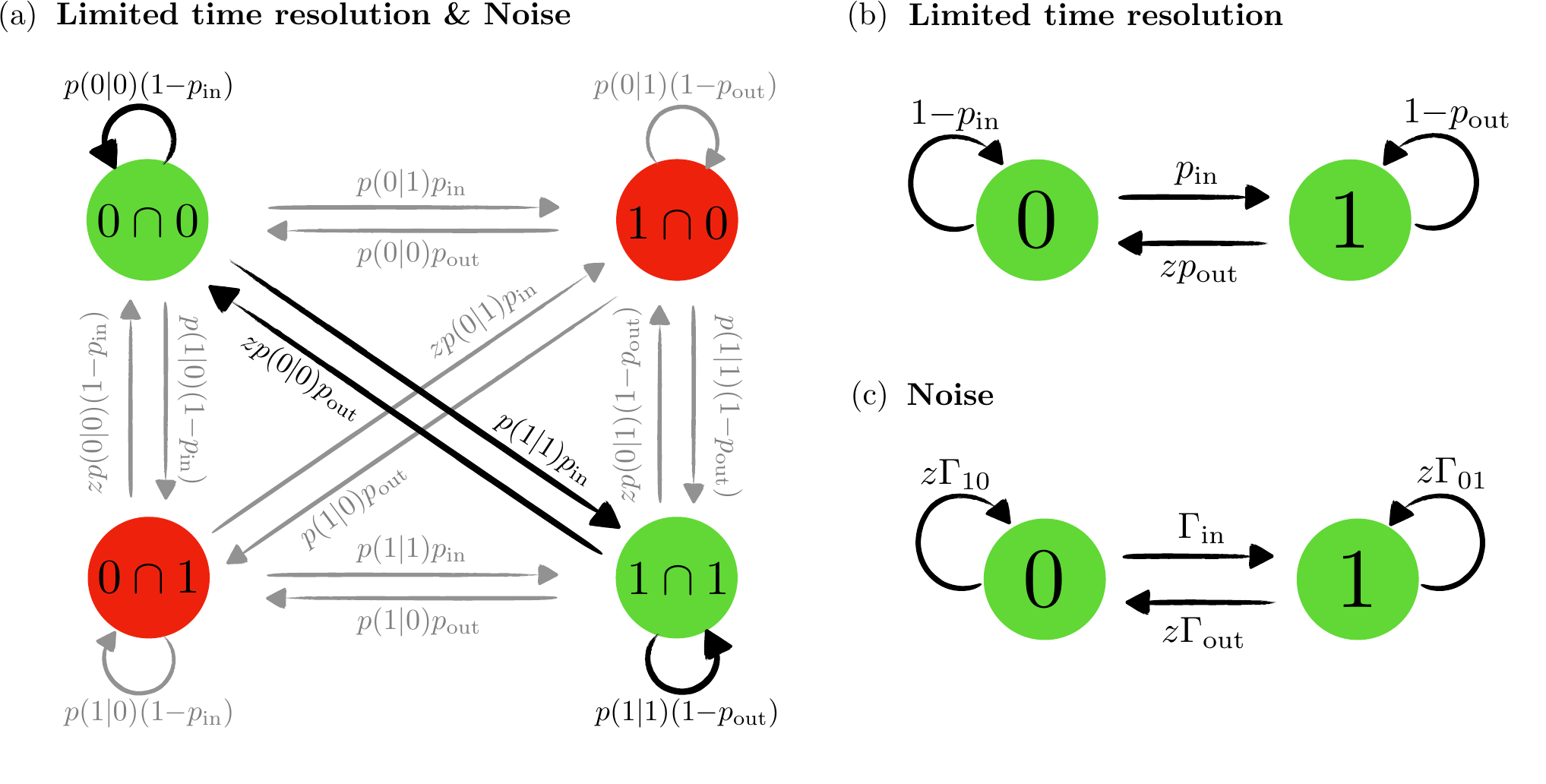}	
        \caption{Modelling electron statistics. (a)~To simulate both a limited time resolution and noise, we apply a four dimensional model. The possible states are indicated via $a\cap b$ denoting that the measurement outcome is $b$ and the true value is $a$. True associations $b=a$ are colored in green and false associations $b\neq a$ are colored in red.  Each time step $\Delta t$, the states are updated due to true tunneling events with transition probabilities $p_\text{in}$ and $p_\text{out}$ and false noise-induced events with probabilities $p(b\vert a)$. Noise-related transitions are indicated as gray arrows such that in the noiseless limit $p(b\vert a)=\delta_{ba}$ only the black arrows survive.  (b)~To simulate only the limited time resolution $\Delta t$, a two-dimensional model is sufficient. In each time step $\Delta t$, an electron may enter/leave the quantum dot with probability $p_\text{in}$/$p_\text{out}$ or remain in the empty/occupied state with probability $1{-}p_\text{in}$/$1{-}p_\text{out}$. Most importantly, in each time step $\Delta t$ at most one tunneling-out event will be counted with the counting variable~$z$. (c)~To simulate only the noise, we add self loops with rates $\Gamma_{10}$ and $\Gamma_{01}$  to the original rate equation. They imitate short fluctuations in the measurement signal of the form $0\rightarrow 1\rightarrow 0$ and $1\rightarrow 0\rightarrow 1$, respectively. In either of those fluctuations, the counter will increase by one, which is covered by multiplying the rates $\Gamma_{10}$ and $\Gamma_{01}$ with the counting variable $z$.  } 
	\label{fig:s4}
\end{figure}
Now, we discuss the \textit{electron statistics} of the quantum-dot system in the parameter regime where it effectively reduces to a two-state system described by the density matrix $\rho\z=(\mel{0}{\rho\z}{0},\mel{1}{\rho\z}{1})$. Hence, the quantum dot is either empty or occupied by one electron. Then, the time evolution is governed by $\dot{\rho}\z={\cal Z} \lbrack{\cal W}\rbrack \rho\z$, see Eq.~\eqref{eq:mastereq_z}, with the generator
\begin{align}
{\cal Z} \lbrack{\cal W}\rbrack =\begin{pmatrix}
-\Gamma_\text{in} & z \Gamma_\text{out}  \\
\Gamma_\text{in} &  -\Gamma_\text{out}  
\end{pmatrix}.
\end{align}
The stationary state is given by $\rho_\text{st}=(\Gamma_\text{out}/\Gamma,\Gamma_\text{in}/\Gamma)$ with $\Gamma=\Gamma_\text{in}{+}\Gamma_\text{out}$. 
Here, we count only electrons leaving the quantum dot, i.e. the counting matrix that enters the operation ${\cal Z}\lbrack...\rbrack$ takes the form
\begin{align}
\Cz =\begin{pmatrix}
(\Cz)_{00} & (\Cz)_{01}  \\
(\Cz)_{10}&  (\Cz)_{11}  
\end{pmatrix}
=
\begin{pmatrix}
1 & z  \\
1&  1  
\end{pmatrix},
\end{align}
where $z$ is the electron counting variable. Thereby, only transitions from $1\rightarrow 0$ increase the counter $N$. 
The system parameters are extracted for an optimal reference measurement with a time resolution of $\Delta t=100\,\mu\text{s}$ and false-count rates as small as $p(1\vert 0)/\Delta t =0.005 \text{kHz}$ and $p(0\vert 1)/\Delta t<10^{-31}\,\text{kHz}$, where we used the threshold $n_\text{th}=13$. 
By evaluating the waiting-time distributions~\cite{brandes_2008} we find the tunneling rates $\Gamma_\text{in}=0.346\,\text{kHz}$ and $\Gamma_\text{out}=0.334\,\text{kHz}$.
Now, by solving the master equation, we can calculate the cumulant generating function $S\qs=\ln \tr \rho\z$  and derive both the ordinary cumulants $\kappa\qs_m$ as well as the factorial cumulants $C\qs_{\text{F},m}$. In the following, we use the results of the optimal reference measurement as a benchmark. Due to the high quality of an optical readout, we refer to the benchmark measurement as the "true dynamics" of the quantum system. By artificially increasing the detection errors, we can clearly distinguish between the true counting statistics of the quantum system and deviations caused by the measurement imperfections.   

\subsection{Measurement errors}
To fully describe the limited time resolution $\Delta t$ and the noise $p(0\vert 1)$ and $p(1\vert 0)$, we use Eq.~\eqref{eq:mastereq_vector} which takes the form
\begin{align}\label{eq:mastereq_error_fourstates}
\begin{pmatrix}
\mel{0}{\rho\z^{(0)}}{0} \\\mel{1}{\rho\z^{(0)}}{1}  \\\mel{0}{\rho\z^{(1)}}{0}  \\ \mel{1}{\rho\z^{(1)}}{1} 
\end{pmatrix}_{t+\Delta t}=
\begin{pmatrix}
p(0\vert 0) & 0 & 0 & 0   \\
0 & p(0 \vert 1) & 0 & 0   \\
0 & 0 & p(1\vert 0) & 0   \\
0 & 0 & 0 & p(1\vert 1)  
\end{pmatrix}
\left[
\begin{pmatrix}
1 & z   \\
1 &  1
\end{pmatrix}
\otimes
\begin{pmatrix}
1-p_\text{in} &  p_\text{out}  \\
p_\text{in} &  1-p_\text{out}  
\end{pmatrix}
\right]
\begin{pmatrix}
\mel{0}{\rho\z^{(0)}}{0} \\\mel{1}{\rho\z^{(0)}}{1}  \\\mel{0}{\rho\z^{(1)}}{0}  \\ \mel{1}{\rho\z^{(1)}}{1} 
\end{pmatrix}_{t},
\end{align}
for the quantum-dot system with the stationary state given by $\boldsymbol{\varrho}_\text{st}=(p(0\vert0)\Gamma_\text{out},p(0\vert1)\Gamma_\text{in},p(1\vert0)\Gamma_\text{out},p(1\vert 1)\Gamma_\text{in})/\Gamma$. The full dynamics takes the form of a discrete Markov chain, see Fig.~\ref{fig:s4}(a). The states $\mel{a}{\rho^{(b)}}{a}=p(a\cap b)$ give the joint probability that we measure $b$ and the true value is $a$. While states with true associations $b=a$ are colored in green, the states with false associations $b\neq a$ are colored in red. 
In each time step $\Delta t$, true tunneling transitions happen with probabilities $p_{\text{in}/\text{out}}=\Gamma_{\text{in}/\text{out}}(1{-}e^{-\Gamma\Delta t })/\Gamma$ and, in addition, noise-induced transitions may occur with probabilities $p(1\vert 0)$ and $p(0\vert 1)$. Noise-related transitions are indicated as gray arrows such that in the noiseless limit $p(b\vert a)=\delta_{ba}$ only the black arrows survive.
We define, in accordance with the notation of the main text, the false-count rate on the bright state as $\Gamma_0^\text{false}=p(1\vert0)/\Delta t$ and on the dark state as $\Gamma_1^\text{false}=p(0\vert1)/\Delta t \ll\Gamma_0^\text{false}$.

By solving Eq.~\eqref{eq:mastereq_error_fourstates}, we find the cumulant generating function via ${S}\sys=\ln \tr\boldsymbol{\varrho}\z (t) $
and can derive the ordinary cumulants ${\kappa}_m\sys$ as well as the factorial cumulants ${C}_{\text{F},m}\sys$ which, now, also account for the systematic errors due to false and missing events. In addition, we apply Eq.~\eqref{eq:obs_msd_mom} to estimate the statistical errors $\Delta \kappa_m\sta$ and $\Delta C_{\text{F},m}\sta$ due to a finite measurement time $T$.
\begin{figure}
        \includegraphics[width=1. \linewidth]{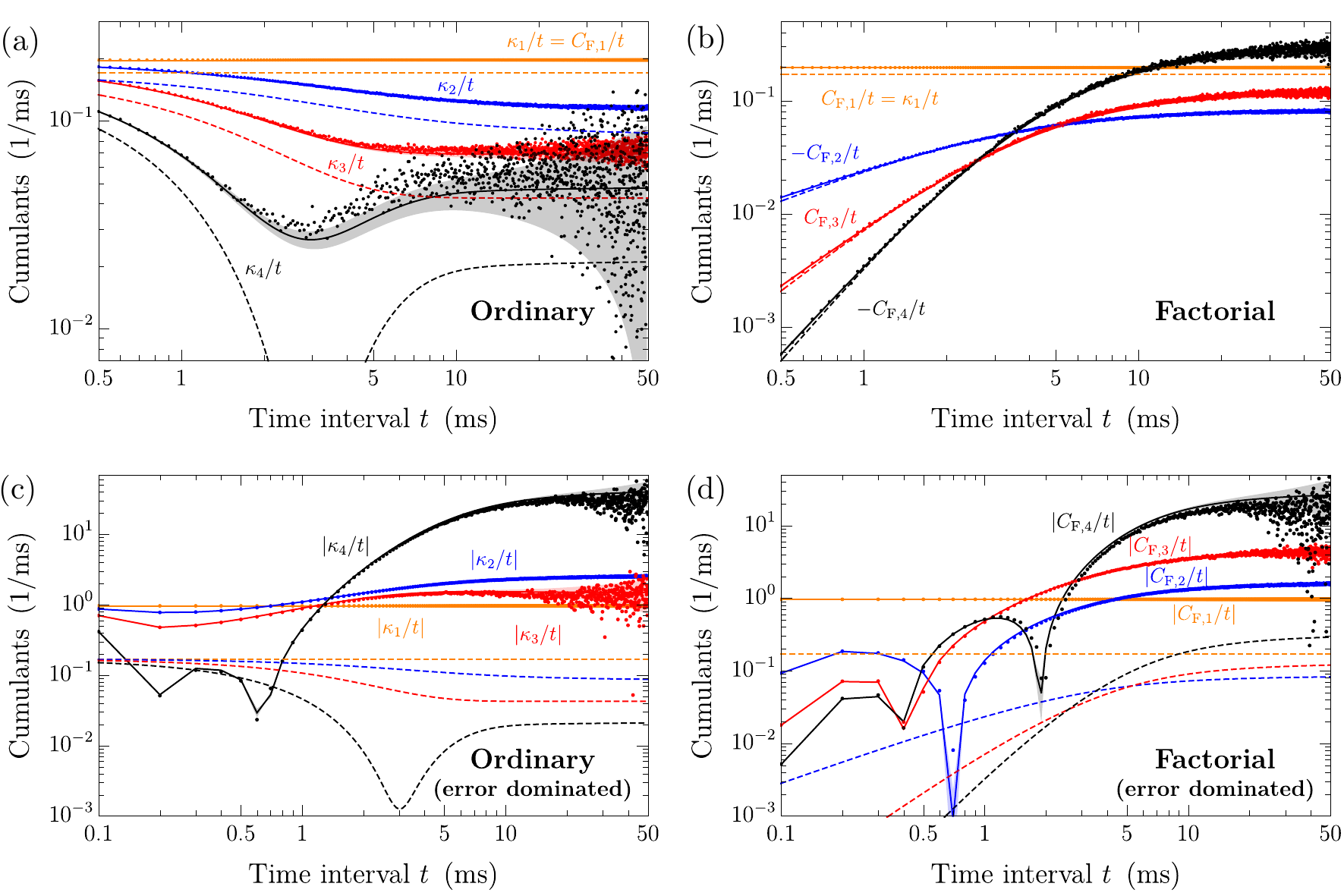}
        \caption{(a)-(d)~Ordinary cumulants $\kappa_m/t$ and factorial cumulants $C_{\text{F},m}/t$ for $m\in\{1,2,3,4\}$ in a double-logarithmic scale as a function of time $t$. Experimental data $\kappa_m\meas/t$ and $C_{\text{F},m}\meas/t$ (dots) is compared with the true statistics of the quantum system  $\kappa_m\qs/t$  and $C_{\text{F},m}\qs/t$ (dashed lines) and with our model $({\kappa}_m\sys\pm \Delta\kappa_m\sta) /t$ and $({C}_{\text{F},m}\sys\pm \Delta C_{\text{F},m}\sta)/t$ (solid lines with continuous error bars) that includes the sources of error due to a limited time resolution $\Delta t$, noise $p(b\vert a)$ as well as a finite measurement time $T$. In (a)-(b), the detector errors are small. The parameters are $\Delta t =50\,\mu\text{s}$, $n_\text{th}=5$, $\Gamma_{0}^\text{false}=  0.059\,\text{kHz}$, $\Gamma_{1}^\text{false}<10^{-12}\,\text{kHz}$, and $T=369~\text{s}$. In (c)-(d), we drastically increased the noise by randomly erasing $95\,\%$ of all detected photons and the parameters are $\Delta t =100\,\mu\text{s}$, $n_\text{th}=0$, $\Gamma_{0}^\text{false}= 2.25\,\text{kHz}$, $\Gamma_{1}^\text{false}=0.014\,\text{kHz}$, and $T=369~\text{s}$. The electron-tunneling rates are $\Gamma_\text{in}=0.346\,\text{kHz}$ and $\Gamma_\text{out}=0.334\,\text{kHz}$.} 
	\label{fig:s5}
\end{figure}

In Fig.~\ref{fig:s5}(a)-(b), we discuss the influence of small measurement errors. 
For ordinary cumulants in Fig.~\ref{fig:s5}(a), the small errors already strongly influence the results and, thus, the experimental cumulants $\kappa_m\meas$ (dots) deviate strongly from the true cumulants $\kappa_m\qs$ (dashed lines) of the quantum system. Only by considering all three sources of error in ${\kappa}_m\sys\pm\Delta \kappa_m\sta$ (solid lines with continuous error bars), we find a very nice agreement with the experiment. 
For factorial cumulants in Fig.~\ref{fig:s5}(b), those same errors are nearly unnoticeable in the measured signal $C_{\text{F},m}\meas$ if $m>1$ and a description with the true factorial cumulants $C_{\text{F},m}\qs$ is completely sufficient. This illustrates the insensitivity of higher-order factorial cumulants to detection errors. 

In Fig.~\ref{fig:s5}(c)-(d), we discuss the case when the measurement errors completely dominate the counting statistics.  
To increase the noise, we randomly erased $95\,\%$ of all detected photons. Together with a small time resolution of $\Delta t =100\,\mu\text{s}$ this leads to a bright-state false-count rate of $\Gamma_{0}^\text{false}=2.25\,\text{kHz}$, much higher than the electron tunneling rates $\Gamma_\text{in}$ and $\Gamma_\text{out}$.
(In contrast, the time resolution $\Delta t=300\,\mu\text{s}$ in Fig.~{\color[rgb]{1,0,0}4} of the main text is much higher, so with $n_\text{th}=0$ the false-count rate $\Gamma_{0}^\text{false}=p(1\vert 0)/\Delta t= 0.038\,\text{kHz}$ remains much smaller there.)
We find that our model ${C}_{\text{F},m}\sys\pm\Delta C_{\text{F},m}\sta$ and ${\kappa}_m\sys\pm\Delta \kappa_m\sta$ (solid lines with continuous error bars) agrees extremely well with the experiment $\kappa_m\meas$ and $C_{\text{F},m}\meas$ (dots). The true cumulants $\kappa_m\qs$ and $C_{\text{F},m}\qs$ (dashed lines) of the quantum system, on the other hand, deviate by orders of magnitude. 
For factorial cumulants $C_{\text{F},m}\meas$, we find in the noise-dominated regime several sign changes which are visible as dips in Fig.~\ref{fig:s5}(d) since the absolute value is presented in a double-logarithmic scale. Although a violation of the sign criterion $(-1)^{m-1}C_{\text{F},m}\meas>0$ is an indication of correlations in the electron statistics~\cite{beenakker_counting_2001,kambly2011factorial,stegmann2015detection,kleinherbers2018revealing}, here, those correlations are merely induced by noise and not by true interactions between the electrons.  
Nevertheless, those sign changes are captured perfectly by our model from Eq.~\eqref{eq:mastereq_error_fourstates}. Hence, even if the dynamics is completely noise-dominated $\Gamma_0^\text{false} \gg \Gamma_\text{in},\Gamma_\text{out}$ our model gives an excellent agreement with the experiment.  

To ensure that the measured telegraph signal still contains information about the dynamics of the quantum system, it must be ensured that both the time resolution is good enough $\Delta t^{-1} > \Gamma_\text{in/out}$, and that noise is sufficiently low, $\Gamma_{0/1}^\text{false}<\Gamma_{\text{in/out}}$. Otherwise, even factorial cumulants cannot correct these errors, cf. Fig.~\ref{fig:s5}(a)-(b) in comparison to Fig.~\ref{fig:s5}(c)-(d).

\begin{figure}
        \includegraphics[width=1.\linewidth]{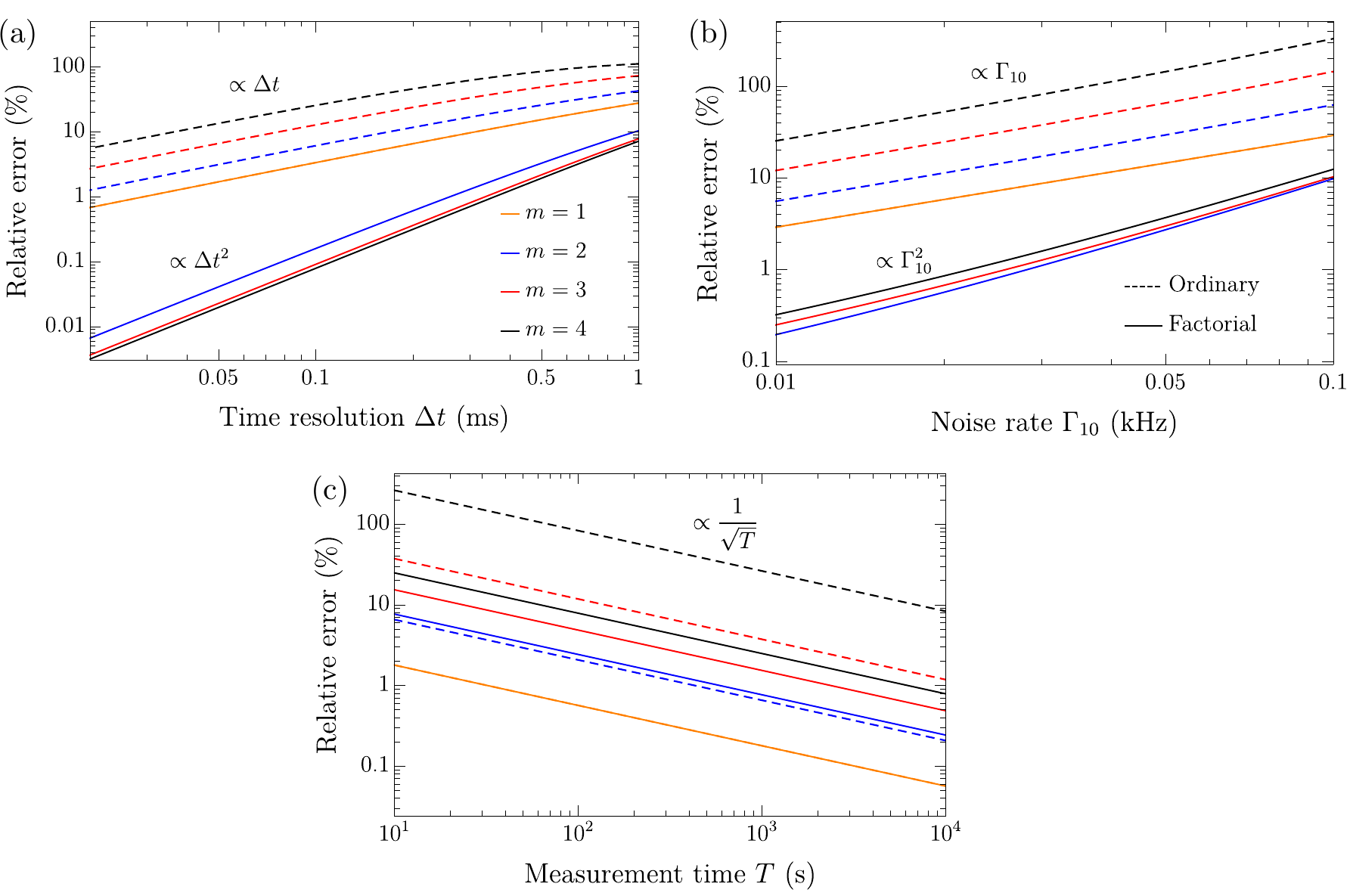}		
        \caption{Relative errors of ordinary and factorial cumulants. (a)-(b)~The relative systematical error of ordinary cumulants $\delta_m=\vert\delta{\kappa}_m\sys\vert/\vert\kappa_m\qs\vert$ (dashed lines) as well as factorial cumulants $\delta_{\text{F},m}=\vert\delta{C}_\text{F,m}\sys\vert/\vert C_\text{F,m}\qs\vert$ (solid lines) is shown as a function of  (a)~the time resolution $\Delta t$ and (b)~the noise rate $\Gamma_{10}$. The relative errors of factorial cumulants $\delta_{\text{F},m}$ lie well below the errors of ordinary cumulants $\delta_{m}$. (c)~The relative statistical error of ordinary cumulants $\delta_m=\Delta \kappa_m\sta/\vert \kappa_m\qs\vert$ (dashed lines) and factorial cumulants $\delta_{\text{F},m}=\Delta C_{\text{F},m}\sta/\vert C_{\text{F},m}\qs\vert$ (solid lines) is shown as a function of the measurement time $T$. The error scales as $1/\sqrt{T}$ according to the law of large numbers. All figures (a)-(c) are evaluated at $t=20\,\text{ms}$ with electron-tunneling rates $\Gamma_\text{in}=0.346\,\text{kHz}$ and $\Gamma_\text{out}=0.334\,\text{kHz}$.} 
	\label{fig:s6}
\end{figure}

\subsubsection{Limited time resolution}
To model only the limited time resolution,  we discretize the master equation $\rho\z(t+\Delta t)={\cal Z} \lbrack e^{{\cal W}\Delta t} \rbrack \rho\z(t)$ in accordance with Eq.~\eqref{eq:mastereq_timeres} and evaluate the propagator as
\begin{align}
{\cal Z} \lbrack e^{{\cal W}\Delta t} \rbrack 
=\begin{pmatrix}
1-p_\text{in} & z p_\text{out}  \\
p_\text{in} &  1-p_\text{out}  
\end{pmatrix}.\label{eq:mastereq_timres_twostates}
\end{align}
The dynamics is  described as a simple Markov chain, see Fig.~\ref{fig:s4}(b). After solving for $\rho\z(t)$, we can evaluate the relative errors of ordinary cumulants $\delta_m=\vert \delta{\kappa}_m\sys\vert/\vert\kappa_m\qs \vert$ and factorial cumulants $\delta_{\text{F},m}=\vert \delta{C}_\text{F,m}\sys \vert/\vert C_\text{F,m}\qs\vert$, which are depicted in Fig.~\ref{fig:s6}(a). For $m=1$, both errors $\delta_1=\delta_{\text{F},1}$ are equal and describe the relative error of the mean value. Remarkably, for $m>1$ the relative errors for factorial cumulants scale as $\delta_{\text{F},m} \propto \Delta t^2$ (solid lines) and are always smaller than the relative errors of ordinary cumulants which scale as $\delta_m \propto \Delta t$ (dashed lines).  
To understand the different scaling behavior, we study the limit of small errors [cf. Sec.~\eqref{sec:smallerrors}] and find that the missing events are described as 
\begin{align}
{\cal W}\z^\text{miss}
=-\frac{\Delta t}{2}\begin{pmatrix}
(z-1)\Gamma_\text{out}\Gamma_\text{in}& 0  \\
0&  (z-1)\Gamma_\text{in}\Gamma_\text{out}  
\end{pmatrix}\propto \mathds{1},
\end{align}
in leading order $\Delta t$. Hence, $\cal{W}^\text{miss}$ describes those missed events, where an electron enters $\Gamma_{\text{in}}$ and leaves $\Gamma_{\text{out}}$ (or vice versa) the quantum dot in a time interval $\Delta t$ unnoticed. This leads to the following error in the counting statistics
\begin{align}
\delta S\sys=-\frac{\Delta t}{2}(z-1) \Gamma_\text{in}\Gamma_\text{out} t +{\cal O}(\Delta t^2),
\end{align}
which holds for all times $t$ because $\cal{W}^\text{miss}\propto \mathds{1}$ .
Since $\delta S\sys(z)$ is linear in the counting variable $z$, the error $\delta_{\text{F},m}$ vanishes in first oder for all factorial cumulants with $m>1$ and starts only in second order $\delta_{\text{F},m}\propto \Delta t^2$. For ordinary cumulants, on the other hand, the first-order error $\delta_m\propto \Delta t$ is present for all orders $m$, because  $\delta S\sys(e^\chi)$ contains powers of arbitrary order in $\chi$. 

\subsubsection{Noise}
To model only the noise, we use  $\dot{\rho}\z=\left( {\cal Z} \lbrack{\cal W}\rbrack + {\cal N}\z \right) \rho\z$, see Eq.~\eqref{eq:mastereq_noise}, and find
\begin{align}
{\cal N}\z=\begin{pmatrix}
\Gamma_{10}(z-1) & 0 \\
0 & \Gamma_{01} (z-1)\end{pmatrix},
\end{align}
where the noise rates are approximated as $\Gamma_{01}\approx \Gamma_1^\text{false}=p(0\vert 1)/\Delta t$ and  $\Gamma_{10}\approx  \Gamma_0^\text{false}=p(1\vert 0)/\Delta t$ for a given $\Delta t$. Hence, the true dynamics is modified by additional self loops, see Fig.~\ref{fig:s4}(c). Those self loops model noise-induced fluctuations in the measured signal of the form $1 \rightarrow 0 \rightarrow 1$ and $0\rightarrow  1\rightarrow 0$ with rates  $\Gamma_{01}$ and $\Gamma_{10}$, respectively. After solving for $\rho\z(t)$, we evaluate the relative errors $\delta_m=\vert \delta{\kappa}_m\sys\vert/\vert\kappa_m\qs\vert$ (dashed lines) and $\delta_{\text{F},m}=\vert\delta{C}_\text{F,m}\sys\vert/\vert C_\text{F,m}\qs\vert$ (solid lines) for $\Gamma_{01}=0$, and find once more that factorial cumulants of order $m>1$ are much less sensitive to errors than ordinary cumulants, see Fig.~\ref{fig:s6}(b). 
By studying the limit of small errors [cf. Sec.~\eqref{sec:smallerrors}], we find that the false events lead to the following contribution to the counting statistics
\begin{align}
\delta S\sys=\frac{\Gamma_{10}}{2} (z-1) t + {\cal O}(\Gamma_{10}^2),
\end{align}
which holds for arbitrary times $t$ if the tunneling rates fulfill $\Gamma_\text{in}\approx\Gamma_\text{out}\approx \Gamma/2$. 
As before, the error  $\delta_{\text{F},m}$ vanishes in first order in $\Gamma_{10}$ for all factorial cumulants with $m>1$ and starts only in second order $\delta_{\text{F},m}\propto   \Gamma_{10}^2$. For ordinary cumulants, on the other hand, the error $\delta_{m}\propto \Gamma_{10}$  starts always in first order. 

\subsubsection{Finite measurement time}
To estimate the statistical error due to a finite measurement time $T$, we use Eq.~\eqref{eq:obs_msd_mom} for  $\Delta \kappa_{m}\sta$ and $\Delta C_{\text{F},m}\sta$, respectively, and replace ${M}_k\sys$ by $M_k$ to eliminate the systematic detection errors due to $\Delta t$ and $\Gamma_{10}$. Then, we define the relative errors as $\delta_m=\Delta \kappa_{m}\sta/\vert\kappa_m\qs\vert$ and $\delta_{\text{F},m}=\Delta C_{\text{F},m}\sta/\vert C_\text{F,m}\qs\vert$. In Fig.~\ref{fig:s6}(c), we see that for all orders $m$, the relative errors of ordinary and factorial cumulants scale as $1/\sqrt{T}$ in accordance with the law of large numbers. Furthermore, we notice that with rising order $m$  the relative error of ordinary cumulants  $\delta_m$ appears to grow faster than the relative error of factorial cumulants $\delta_{\text{F},m}$. This can be understood by inspecting the long-time limit $\Gamma t\gg1$ for symmetric rates $\Gamma_\text{in}\approx\Gamma_\text{out}\approx \Gamma/2$, where we find $\kappa_m\approx (1/2)^m \Gamma t/2$ and $C_{\text{F},m}\approx(1/2)^{(m)}  \Gamma t/2$. Since for long times the absolute errors are similar $\Delta \kappa_{m}\sta\approx \Delta C_{\text{F},m}\sta$, cf. Eq.~\eqref{eq:cum_msd}, the ratio between the relative errors for $m>2$ is given by
\begin{align}
\frac{\delta_{m}}{\delta_{\text{F},m}}=\frac{\Delta \kappa_{m}\sta/\kappa_m\qs}{\Delta C_{\text{F},m}\sta/C_\text{F,m}\qs}\approx \frac{C_{\text{F},m}\qs}{\kappa_m\qs}\approx \frac{(1/2)^{(m)}}{(1/2)^{m}}=(2m-3)!! \gg 1.
\end{align}
Thus, the relative error of ordinary cumulants $\delta_m$ is larger than the relative error of factorial cumulants $\delta_{\text{F},m}$ by the double factorial $(2m-3)!!$. 
By defining the times $t_m^*$ and $t^*_{\text{F},m}$ when the relative errors are as big as $\delta_m=1$ and $\delta_{\text{F},m}=1$, we find using Eq.~\eqref{eq:cum_msd} that $t^*_{\text{F},m}/t^*_m=\lbrack (2m-3)!! \rbrack^{2/(m-1)}\gg1$ for $m>2$. As a consequence, higher-order factorial cumulants can be resolved for longer times $t$ than ordinary cumulants.

\addcontentsline{toc}{part}{References}
\stoptocentries